\documentclass[pra,onecolumn,preprintnumbers,amsmath,amssymb,nofootinbib]{revtex4}
\pdfoutput=1 
\usepackage{hyperref}
\usepackage{kbordermatrix}
\usepackage{mdframed}
\usepackage{braket}
\usepackage{xy}
\usepackage{qcircuit}
\usepackage{float}
\usepackage{graphicx}

\DeclareMathOperator{\Tr}{Tr}

\begin{document}

\title{Quantum machine learning for data scientists}
\author{Dawid Kopczyk}
\affiliation{Quantee Limited, Manchester, United Kingdom}

\begin{abstract}
This text aims to present and explain quantum machine learning algorithms to a data scientist in an accessible and consistent way. The algorithms and equations presented are not written in rigorous mathematical fashion, instead, the pressure is put on examples and step by step explanation of difficult topics. This contribution gives an overview of selected quantum machine learning algorithms, however there is also a method of scores extraction for quantum PCA algorithm proposed as well as a new cost function in feed-forward quantum neural networks is introduced. The text is divided into four parts: the first part explains the basic quantum theory, then quantum computation and quantum computer architecture are explained in section two. The third part presents quantum algorithms which will be used as subroutines in quantum machine learning algorithms. Finally, the fourth section describes quantum machine learning algorithms with the use of knowledge accumulated in previous parts. 
\end{abstract}

\maketitle

\tableofcontents 

\section{Introduction}
Machine learning is part of computer science area which aims to recognize patterns and learn from data in order to output correct predictions. It could be considered as a form of artificial intelligence supporting government analysis, medical reports, business decisions, financial risk management and other areas where decisions and optimization are based on information stored digitally. Due to an increasing amount of data stored by the companies across the world \cite{MH} and several breakthroughs in working software, machine learning is increasingly important in industry. In the last couple of years, researchers have been investigating whether quantum computers can be used to improve the performance of machine learning algorithms. A quantum computer takes advantage of quantum mechanical effects such as superposition and entanglement to solve a certain set of problems faster than classical computers. Even though quantum computers are still at the experimental stage (with some major breakthrough made by IBM that announced to build 16 qubits processor \cite{IBM} with computation resources available in the cloud), the quantum algorithms have been developed for the last two decades. The quantum algorithms involve problems such as factorization of large numbers and optimization with the latter effectively used in a speed-up of machine learning algorithms. Quantum machine learning is a relatively new area of study with the recent work on quantum versions of supervised and unsupervised algorithms. The major difficulty for a non-physicist person such as a data scientist is the requirement of quantum physics theory and scientific notation knowledge. It can create some cognitive barriers to understand benefits and limitations of the quantum algorithms. Most of articles are written for a quantum physicist, leaving little or no explanation of remarkable techniques and discoveries of the quantum algorithms and the quantum machine learning algorithms. This text aims to present and explain the quantum machine learning algorithms to a data scientist in an accessible and consistent way as well as provides an introduction for a physicist interested in the topic. Moreover, it is a review of circuit-based quantum machine learning algorithms with their benefits and limitations listed. In order to properly explain quantum machine learning to a non-physicists, the paper presents an absolute minimum of quantum theory required to understand quantum computation. Then, the quantum algorithms and the quantum machine learning algorithms are presented in a step by step manner with accessible examples. Hopefully, this will allow an ambitious data scientist to understand the possibilities and limitations associated with the use of quantum computers in machine learning as well as gain knowledge about the mathematics and logic behind qunatum machine learning algorithms. 

\section{Basic quantum theory}
This section summarizes basic concepts of quantum theory required to understand how quantum algorithms work and how quantum effects can be used in the machine learning algorithms. Quantum mechanics seems to be counter-intuitive due to the fact that the classical world observed by our senses is too macroscopic in order to notice the rules of the microscopic world. The brief introduction to quantum theory aims to explain the quantum realm to a data scientist in a sufficient way to understand how it is used to speed-up machine learning algorithms. As a good starter, the beginnings of quantum mechanics are explained. \par
Before the twentieth century, physicists used to hold the view that the matter surrounding us could be either made up of tiny particles or consists of waves. The light was viewed as an electromagnetic wave that \textit{just like} a surface water wave or a sound wave can interfere with itself making the wave amplitude greater or lower. On the other hand, the matter was attributed to be made up of particles. This view was falsified by series \cite{Broglie,Davisson,EinsteinPhoto} of experiments, among others the photoelectric effect. The photoelectric effect assumes that electrons can be ejected from the surface of metal plate when light shines on it. According to classical electromagnetic theory increasing light amplitude would increase the kinetic energy of emitted photoelectrons, while increasing the frequency would increase measured current. Interestingly, this was contrary to experimental observations. The effect was explained by Albert Einstein \cite{EinsteinPhoto} as he assumed that the light is a collection of particles called photons and it earned him the Nobel Prize in 1921. As it turns out later all matter in the universe manifests both particle-like and wave-like behavior, which provided the basis for a new exciting area of human knowledge, namely quantum mechanics. This is a somewhat counter-intuitive approach to the reality, so in the next sections, we will try to explain it with a big help of mathematics.

\subsection{Quantum states}
Imagine a set of positions $\{x_1.x_2,\dots,x_n\}$ in which a particle can be detected:
\begin{figure}[H]
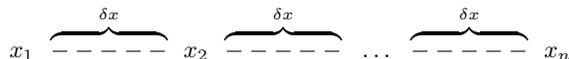

\begin{align*}
x_1 \phantom{0} \overbrace{ ----- }^{\delta x} \phantom{0} x_2 \phantom{0} \overbrace{ ----- }^{\delta x} \phantom{0} \dots \phantom{0} \overbrace{ ----- }^{\delta x} \phantom{0} x_n
\end{align*}
\vspace*{-1em}
\caption{Discrete set of positions in which a particle can be detected.} 
\label{fig:position}
\end{figure} \noindent
The gap between the positions $\delta x$ is very small to provide good approximation of continuous line. The current set-up can be represented by states corresponding to a particle being detected at given locations. The state corresponding to a particle being detected at location $x_1$ could be written as a column vector:
\begin{align}
[1, 0, \dots, 0]^T \,. 
\end{align}
The next state corresponding to a particle being detected at location $x_2$ is associated with a column vector which is orthogonal to the previous one:
\begin{align}
[0, 1, \dots, 0]^T \,.  
\end{align}
Following the logic we would expect that each of these states would be an unit vector $e_i$ in the standard basis. In terms of classical mechanics the current description is complete. However, the quantum mechanics is not so boring. In quantum realm this set-up is described by a quantum state. The quantum states are denoted using Dirac notation $\ket{a}$ which is equivalent to a $n$ dimensional vector with complex numbers as entries. Although, we can use a standard vector notation, it is just easier to follow what other physicists do. The state corresponding to a particle being detected at position $x_i$ is denoted as $\ket{x_i}$. The essence of quantum mechanics is that a particle before detection could be at state which is a mixture of states $\{\ket{x_1},\ket{x_2},\dots,\ket{x_n}\}$. Mathematically, the mixture is denoted by an arbitrary state $\ket{\psi}$ which is a linear combination of the basis states $\{\ket{x_1},\ket{x_2},\dots,\ket{x_n}\}$ weighted by complex amplitudes $\{c_1,c_2,\dots,c_n\}$:
\begin{align}
\ket{\psi} &= c_1\ket{x_1}+c_2\ket{x_2}+\dots+c_n\ket{x_n} \nonumber \\
&= c_1\begin{bmatrix}1 \\ 0 \\ \vdots \\ 0 \end{bmatrix}+c_2\begin{bmatrix}0 \\ 1 \\ \vdots \\ 0 \end{bmatrix}+\dots+c_n\begin{bmatrix}0 \\ 0 \\ \vdots \\ 1 \end{bmatrix}=\begin{bmatrix}c_1 \\ c_2 \\ \vdots \\ c_n \end{bmatrix} \,.
\end{align}
It is simple as it is: quantum states are denoted using ket notation $\ket{a}$, which is a column vector in $\mathbb{C}^n$ complex space (quantum states are also defined on a space with infinite number of dimensions, but this is not relevant to the quantum algorithms presented in the text). However, several question arise:
\begin{itemize}
\item What does it exactly mean that the state of a particle is a linear combination?
\item Why we use complex numbers as weights?
\end{itemize}
We will try to address and answer the following questions, revealing the beauty of quantum mechanics. The quantum state being in a linear combination is equivalent to say that a particle is in a quantum superposition of the basis states. The state $\ket{\psi}$ means that a particle before measurement is \textit{a probability wave} corresponding to the probabilities of being measured at $\{x_1,x_2,\dots,x_n\}$ locations. Thus, a question \textit{where is a particle before measurement?} according to Copenhagen interpretation of quantum mechanics is meaningless. In macroscopic terms matter seems to be localized and we could not see an item being in many places at the same time. This is due to the fact that unbelievably large number of particles is interacting with themselves so that the information from quantum states is transferred to the environment, making the quantum behavior \textit{lost} in the jungle of interacting particles. This process is called decoherence and results in leaking of quantum nature of system to the environment. Thus, we cannot see something being \textit{before measurement}, as even our brains are part of this interacting system. In quantum scales when we analyze behavior of a single or a few particles in isolation, the quantum effects arise. The nature of quantum mechanics is confirmed by large amount of experiments, so that this fact needs to be believed and incorporated. The considerations on the border of physics and philosophy about what happens with particle before measurement, although very interesting, are not required to understand quantum machine learning algorithms (see Bohr-Einstein debates in Ref.~\cite{Bohr}). For us it is sufficient to say that a particle before measurement consists of different probabilities corresponding to different outcomes obtained after measurement. \par
Returning to the quantum state $\ket{\psi}$ note that there are many possible superpositions controlled by the value of weights $\{c_1,c_2,\dots,c_n\}$. These weights are called probability amplitudes and are strictly connected to what happens with a particle after measurement. The norm square of complex number $|c_i|^2$ gives us the probability of findinga particle in the state $\ket{x_i}$ after measurement. Due to the fact that we deal with probabilities, the probability amplitudes should be properly normalized so that:
\begin{align}
|c_1|^2+|c_2|^2+\dots+|c_n|^2=\sum_{i=1}^n |c_i|^2 = 1 \,.
\end{align}
As an example we have an arbitrary state:
\begin{align}
\ket{\psi} = c_1\ket{x_1}+c_2\ket{x_2}=\frac{i}{2}\begin{bmatrix} 1 \\[2pt] 0 \end{bmatrix} + \frac{\sqrt{3}}{2}\begin{bmatrix} 0 \\[2pt] 1 \end{bmatrix} = \begin{bmatrix} \frac{i}{2} \\[2pt] \frac{\sqrt{3}}{2} \end{bmatrix} \,.
\end{align}
What is the probability that a particle after measurement will be in the state $\ket{x_1}$? The answer implies taking norm square of $c_1$:
\begin{align}
|c_1|^2=c_1^*c_1=\frac{-i}{2}\frac{i}{2}=\frac{-i^2}{2^2}=\frac{1}{4} \,.
\end{align}
This can be also calculated in a more systematic way. We claim that the probability of finding a particle $\ket{\psi}$ in a state $\ket{\phi}$ after measurement is expressed as:
\begin{align}
|\braket{\phi \vert \psi}|^2
\end{align}
where bra notation $\bra{a}$ represents a conjugate transpose of ket $\ket{a}$. The bracket $\braket{a\vert a}$ is an inner product of two vectors, which is just a number. To see how it works we calculate the inner product from the example:
\begin{align}
\braket{x_1\vert \psi} &= \bra{x_1}c_1\ket{x_1}+\bra{x_1}c_2\ket{x_2}=c_1\braket{x_1 \vert x_1}+c_2\braket{x_1 \vert x_2} \nonumber \\
&= \frac{i}{2}\begin{bmatrix} 1 & 0 \end{bmatrix}\begin{bmatrix} 1 \\[2pt] 0 \end{bmatrix} + \frac{\sqrt{3}}{2}\begin{bmatrix} 1 & 0 \end{bmatrix}\begin{bmatrix} 0 \\[2pt] 1 \end{bmatrix} = \frac{i}{2}
\end{align}
and then the norm square is:
\begin{align}
|\braket{\phi \vert \psi}|^2=\braket{\phi \vert \psi}^*\braket{\phi \vert \psi}=\frac{-i}{2}\frac{i}{2}=\frac{1}{4}
\end{align}
yielding the expected result. The orthogonality of the basis states could be interpreted in terms of probability. The probability of measuring the particle $\ket{x_1}$ to be in the state $\ket{x_1}$ always yields $|\braket{x_1 \vert x_1}|^2=1$, whereas we would never find the same particle being in the state $\ket{x_2}$ as $|\braket{x_2 \vert x_1}|^2=0$. \par
We have stated that particles manifests wave-like behavior and promised to explain that fact mathematically. The wave-like interference is fully explained by the presence of complex numbers in probability amplitudes. The probabilities in real numbers when added are always greater or equal: $p_1+p_2 \geq p_1$ and $p_1+p_2 \geq p_2$. The complex amplitudes when squared are also real, but now the addition of complex numbers $|c_1+c_2|^2$ can increase or decrease the probability. The probability amplitude $c_1=\frac{i}{2}$ when squared is equal to probability $|c_1|^2=\frac{1}{4}$. The probability amplitude $c_2=\frac{-i}{2}$ when squared is also equal to probability $|c_2|^2=\frac{1}{4}$, however the sum of probability amplitudes $c_1+c_2$ yields probability $|c_1+c_2|^2=|\frac{-i+i}{2}|^2=0$ which is certainly lower. The complex numbers can cancel or overlay each other, which has a physical meaning of interference. This is the core of quantum mechanics allowing to explain wave-like behavior of particles. 

\subsection{Quantum observables}
The physics is all about measuring and analyzing quantities such as position, momentum or energy. These quantities are called observables and can be retrieved from the current state of a system. In classical physics observable $F$ is a function that takes state $S$ and outputs real number $x$ which corresponds to measured quantity, that is:
\begin{align}
F(S)=x \,.
\end{align}
As an example consider we measure the observable which is the heat of the gas and want to output the energy $E$ of the system. The state $S$ is characterized by the temperature of the system $T$ and there exists a real-valued function $F$ that allows us to calculate energy, so that $F(T)=E$. In quantum physics an observable is not a real-valued function, but is represented by a matrix $O$ that acts on a quantum state $\ket{\psi}$. Just like the function $F$ in classical physics, the matrix $O$ allows us to retrieve quantity from a system, however in quantum physics the result of measurement is discretized. The eigenvalues $\lambda_i$ of the matrix $O$ are the only possible values observable can take after being measured. The eigenvectors $\ket{a_i}$ can be interpreted as states in which the system is left after measuring the associated eigenvalue $\lambda_i$. This is written as:
\begin{align}
O\ket{\psi} \rightarrow \lambda_i\ket{a_i}
\end{align}
with the arrow representing measurement and collapse of the state superposition $\ket{\psi}$ to the state $\ket{a_i}$. Each quantity we want to retrieve from quantum state is associated with a different observable. In case we would like to measure position and momentum having state $\ket{\psi}$, the observables corresponding to the position and momentum are represented by different matrices. Each of these matrices has eigenvalues that are the only possible values of measured quantity, that is either position or momentum. As an example of observable, we take the position from the set-up presented in Fig.~(\ref{fig:position}). A value measured is the position of a particle on the line. We know that each measurement can yield position $x_i$ with different probabilities $|c_i|^2$. The state after the measurement is one of the $\ket{x_i}$ basis states. As a result we are looking for a matrix which eigenvalues are the real numbers $x_i$ and associated eigenvectors are exactly $\ket{x_i}$:
\begin{align}
O\ket{\psi} = \sum_{i=1}^n x_i c_i \ket{x_i} \rightarrow x_i\ket{x_i} \,.
\end{align}
Due to the fact that the eigenvalues of the matrix must be real (we do not observe in the universe position or momentum equal $1+i$), $O$ must be a Hermitian matrix. The searched matrix has a form of:
\begin{align}
O = \begin{bmatrix} x_1 & 0 & \cdots & 0 \\ 0 & x_2 & \cdots & 0 \\ \vdots & \vdots & & \vdots \\ 0 & 0 & \cdots & x_n  \end{bmatrix} \,.
\end{align}
The matrix is Hermitian, the eigenvalues are obviously $x_i$ and the corresponding eigenvectors are $\ket{x_i}$. \par
The distribution of possible outcomes $\lambda_i$ is governed by probabilities $ |c_i|^2$. Making multiple measurements of a particle in the same state $\ket{\psi}$ we could be interested in an expectation value of the observable $O$. Note that the phrase \textit{multiple measurements} means that the state is prepared, measured and then prepared from scratch again. In case we measure the same state immediately after previous measurement we would simply get collapsed state with 100\% probability. An expectation value of the observable $O$ is denoted as $\langle O \rangle$ and can be calculated as follows:
\begin{align}
\langle O \rangle &= \braket{\psi \vert O \vert \psi} \nonumber \\
&= (c_1^*\bra{a_1}+c_2^*\bra{a_2}+\dots+c_n^*\bra{a_n}) (c_1O\ket{a_1}+c_2O\ket{a_2}+\dots+c_n O\ket{a_n}) \nonumber \\
&= (c_1^*\bra{a_1}+c_2^*\bra{a_2}+\dots+c_n^*\bra{a_n}) (c_1\lambda_1\ket{a_1}+c_2\lambda_2\ket{a_2}+\dots+c_n \lambda_n\ket{a_n}) \nonumber \\
&= \lambda_1|c_1|^2+\lambda_2|c_2|^2+\dots+\lambda_n|c_n|^2 \label{eq:expectation}
\end{align}
with the last line in Eq.~(\ref{eq:expectation}) being a statistical definition of expectation value:
\begin{align}
E(\lambda)=\sum_{i=1}^n \lambda_i p_i \,.
\end{align}
In these calculations we have used the fact that the basis states $\ket{a_i}$ and $\ket{a_j}$ are orthogonal for $i\neq j$ meaning that $\braket{a_j \vert a_i}=0$ and the basis states are actually the eigenvectors of matrix $O$ meaning that $O\ket{a_i}=\lambda_i\ket{a_i}$. As an example we calculate an expectation value of position for state:
\begin{align}
\ket{\psi} = \frac{i}{2}\begin{bmatrix} 1 \\[2pt] 0 \end{bmatrix} + \frac{\sqrt{3}}{2}\begin{bmatrix} 0 \\[2pt] 1 \end{bmatrix} = \begin{bmatrix} \frac{i}{2} \\[2pt] \frac{\sqrt{3}}{2} \end{bmatrix} \,.
\end{align}
The possible positions measured are $x_1=1$ and $x_2=2$ so that the observable is represented by matrix:
\begin{align}
O = \begin{bmatrix} 1 & 0 \\ 0 & 2 \end{bmatrix} \,.
\end{align}
The expectation value $\langle O \rangle$ is calculated in a following way:
\begin{align}
\langle O \rangle &= \braket{\psi \vert O \vert \psi} \nonumber \\
&=\begin{bmatrix} \frac{-i}{2} & \frac{\sqrt{3}}{2} \end{bmatrix} \begin{bmatrix} 1 & 0 \\[2pt] 0 & 2 \end{bmatrix} \begin{bmatrix} \frac{i}{2} \\[2pt] \frac{\sqrt{3}}{2} \end{bmatrix} \nonumber  \\
&= 1*\frac{1}{4}+2*\frac{3}{4} = \frac{7}{4} = 1.75 \,.
\label{eq:expectationO}
\end{align}
We might be also interested in the variance of outcomes, i.e. the spread of the possible results around the expectation value. The variance $Var(O)$ is defined as:
\begin{align}
Var(O) &= \braket{\psi \vert (O - \langle O \rangle)^2 \vert \psi} \nonumber  \\
&= (c_1^*\bra{a_1}+\dots+c_n^*\bra{a_n}) (c_1(O - \langle O \rangle)^2\ket{a_1}+\dots+c_n (O - \langle O \rangle)^2\ket{a_n}) \nonumber \\
&= (c_1^*\bra{a_1}+\dots+c_n^*\bra{a_n}) (c_1(\lambda_1 - \langle O \rangle)^2\ket{a_1}+\dots+c_n (\lambda_n - \langle O \rangle)^2\ket{a_n}) \nonumber \\
&= (\lambda_1 - \langle O \rangle)^2|c_1|^2+\dots+(\lambda_n - \langle O \rangle)^2|c_n|^2
\end{align}
which is a statistical definition of variance:
\begin{align}
Var(\lambda)=\sum_{i=1}^n (\lambda_i-\bar{\lambda})^2 p_i
\end{align}
with $\bar{\lambda}$ denoting mean of the data. The variance of the observable $O$ acting on the state $\ket{\psi}$ from the previous example in which $\langle O \rangle = 1.75$ is calculated as follows:
\begin{align}
Var(O) &= \braket{\psi \vert (O - \langle O \rangle)^2 \vert \psi} \nonumber \\
&=\begin{bmatrix} \frac{-i}{2} & \frac{\sqrt{3}}{2} \end{bmatrix} \begin{bmatrix} (1-1.75)^2 & 0 \\[2pt] 0 & (2-1.75)^2 \end{bmatrix} \begin{bmatrix} \frac{i}{2} \\[2pt] \frac{\sqrt{3}}{2} \end{bmatrix} \nonumber \\
&= 0.5625*\frac{1}{4}+0.0625*\frac{3}{4} = 0.1875
\end{align}
resulting in standard deviation of approximately $\sqrt{0.1875} \approx 0.433$. 

\subsection{Measurement}
The previous section provided some intuition and basic facts about measurement in quantum mechanics. This section is to summarize the topic and provide the comparison to classical measurement. In classical physics the measurement is characterized by two assumptions:
\begin{itemize}
\item Theoretically, the measurement leaves the system in the same state as it was before,
\item The result of measurement is predictable. It means that if the experiment is to be repeated, we would anticipate exactly the same outcome. 
\end{itemize}
In quantum scales, these assumptions turned out to be wrong. The measurement in quantum mechanics is:
\begin{itemize}
\item Irreversible operation that transforms the general state $\ket{\psi}$ into an eigenvector $\ket{a_i}$ of measured observable $O$. It is also said that the state $\ket{\psi}$ has collapsed to the eigenvector $\ket{a_i}$,
\item The result of measurement is uncertain and is always one of the eigenvalue $\lambda_i$ of measured observable $O$. The eigenvalue will be measured with probability $|\braket{a_i \vert \psi}|^2$, where $\ket{a_i}$ is the eigenvector corresponding to eigenvalue $\lambda_i$.
\end{itemize}
Two facts emerge from the points listed. One is that if we perform measurement immediately after the first measurement we would observe state $\ket{a_i}$ with $100\%$ probability. It is caused by the fact that the system has already collapsed to one of its eigenvectors and measurement of eigenvector always yields the same eigenvector. The second fact is related to the order of different observables measurement. The measurement is an irreversible operation, thus performing measurement of first observable will have an impact on the measurement of second observable immediately after the first measurement. \par
In quantum physics, there exists intriguing fact about the measurement of two observables. This fact is a well-known Heisenberg uncertainty principle which states that product of the variances of two observables is always greater than a threshold equal to one-fourth of squared expected value of their commutator:
\begin{align}
Var(O_1)Var(O_2) \geq \frac{1}{4}\left|\langle [O_1,O_2] \rangle \right|^2 \,.
\end{align}
The commutator is defined as a difference:
\begin{align}
[O_1,O_2]=O_1O_2-O_2O_1
\end{align}
which for matrices is not necessarily zero. In order to fully understand the Heisenberg uncertainty principle we analyze the following example:
\begin{itemize}
\item Prepare $k$ repetitions of experiment (exactly the same states are measured in each experiment),
\item Measure position and then momentum in each experiment,
\item Write the results and calculate the variance for both position and momentum outcomes.
\end{itemize}
It turns out that the product of these variances will be always greater than a non-zero threshold because the Hermitian matrices corresponding to position and momentum do not commute $[O_1,O_2] \neq 0$. The exact value of the threshold is $\frac{\hbar^2}{4}$ where $\hbar$ is a reduced Planck's constant, however the numeric value is not required to understand the principle. The non-commutativity is interpreted as the fact that observables $O_1$ and $O_2$ do not share the eigenvectors. In case $O_1$ and $O_2$ commute they have identical eigenvectors $\ket{a_i}$ corresponding to some eigenvalues $\lambda_i$ and $\mu_i$ so that:
\begin{align}
\begin{cases}
O_1\ket{a_i} &= \lambda_i\ket{a_i} \\
O_2\ket{a_i} &= \mu_i\ket{a_i} \,.
\end{cases}
\end{align}
Thus the second observable $O_2$ can be measured without disturbing already collapsed eigenvector $\ket{a_i}$. Some of the sources explain that the product of variances is greater than some threshold for non-commuting observables because the measurement itself disturbs the state. Although the effects of disturbing are not negligible, this is not the true reason why the Heisenberg uncertainty principle holds. The principle arises from the fact that particles are waves, not point objects and it is a fundamental property of the universe, so that even if the measurement is done without any disturbance of the system (obviously except collapsing of state) the uncertainty for non-commuting observables still exists.

\subsection{Assembling quantum states}
Assume we want to analyze multi-particle states instead of only one particle states. The machinery that should be used to accomplish that is called tensor product of state spaces and the procedure is called assembling of quantum states. Having $k$ independent particle states: $\{\ket{\psi_1}, \ket{\psi_2}, \dots, \ket{\psi_k}\}$ we can describe them by general state $\Psi$:
\begin{align}
\ket{\Psi}=\ket{\psi_1} \otimes \ket{\psi_2} \otimes \dots \otimes \ket{\psi_k}=\ket{\psi_1\psi_2\dots\psi_k} \,.
\end{align}
Thus, if $\ket{\psi_i}$ are $n$-dimensional vectors for each $i$, then the state $\ket{\Psi}$ will have $n^k$ dimensions. As an example, let us add another particle to the example presented in Fig.~(\ref{fig:position}), so that the system is illustrated by:
\begin{figure}[H]
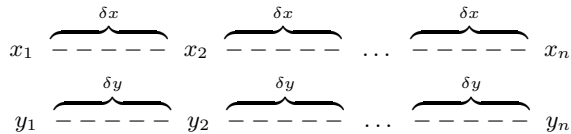

\begin{align*}
x_1 \phantom{0} \overbrace{ ----- }^{\delta x} \phantom{0} x_2 \phantom{0} \overbrace{ ----- }^{\delta x} \phantom{0} \dots \phantom{0} \overbrace{ ----- }^{\delta x} \phantom{0} x_n \\
y_1 \phantom{0} \overbrace{ ----- }^{\delta y} \phantom{0} y_2 \phantom{0} \overbrace{ ----- }^{\delta y} \phantom{0} \dots \phantom{0} \overbrace{ ----- }^{\delta y} \phantom{0} y_n
\end{align*}
\vspace*{-1em}
\caption{Two discrete sets of positions in which particle $x$ and $y$ can be detected.} 
\end{figure} \noindent
Now, the possible states are the combinations of $x$ particle position and $y$ particle position. The general state $\ket{\Psi}$ for $n=2$ possible outcomes is constructed as follows:
\begin{align}
\ket{\Psi}&=\ket{\psi_x} \otimes \ket{\psi_y} \nonumber \\
&= \left(c_{1,x}\ket{x_1}+c_{2,x}\ket{x_2}\right) \otimes \left(c_{1,y}\ket{y_1}+c_{2,y}\ket{y_2}\right) \nonumber \\
&= \overbrace{c_{1,x}c_{1,y}}^{c_1}\ket{x_1y_1}+\overbrace{c_{1,x}c_{2,y}}^{c_2}\ket{x_1y_2}+\overbrace{c_{2,x}c_{1,y}}^{c_3}\ket{x_2 y_1}+\overbrace{c_{2,x}c_{2,y}}^{c_4}\ket{x_2 y_2} \nonumber \\
&= c_1 \begin{bmatrix} 1 \\ 0 \end{bmatrix} \otimes \begin{bmatrix} 1 \\ 0 \end{bmatrix} + c_2 \begin{bmatrix} 1 \\ 0 \end{bmatrix} \otimes \begin{bmatrix} 0 \\ 1 \end{bmatrix} + c_3 \begin{bmatrix} 0 \\ 1 \end{bmatrix} \otimes \begin{bmatrix} 1 \\ 0 \end{bmatrix} + c_4 \begin{bmatrix} 0 \\ 1 \end{bmatrix} \otimes \begin{bmatrix} 0 \\ 1 \end{bmatrix} \nonumber \\ 
&= c_1 \begin{bmatrix} 1 \\ 0 \\ 0 \\ 0 \end{bmatrix} + c_2 \begin{bmatrix} 0 \\ 1 \\ 0 \\ 0 \end{bmatrix} + c_3 \begin{bmatrix} 0 \\ 0 \\ 1 \\ 0 \end{bmatrix} + c_4 \begin{bmatrix} 0 \\ 0 \\ 0 \\ 1 \end{bmatrix}
\end{align} 
where we implicitly assume that the outcomes of particle measurement are described by eigenvectors $e_1=[1,0]^T$ and $e_2=[0,1]^T$. The state $\ket{\Psi}$ is $2^2=4$ dimensional vector and is in a superposition of all possible position measurement outcomes of two particles. The interpretation of complex numbers $c_i$ is exactly the same as in one particle states, for instance $c_2$ is a probability amplitude for measuring the first particle at position $x_1$ and the second particle at position $y_2$. \par
The assembling of quantum states is strictly related to another astonishing property of quantum theory, that is quantum entanglement. This concept will be explained based on a simple example. Assume existence of two quantum states:
\begin{align}
\ket{\psi}=\frac{1}{2}\ket{aa}+\frac{1}{2}\ket{ba}+\frac{1}{2}\ket{ab}+\frac{1}{2}\ket{bb}
\end{align}
and
\begin{align}
\ket{\phi}=\frac{1}{\sqrt{2}}\ket{aa}+\frac{1}{\sqrt{2}}\ket{bb} \,.
\end{align}
Both states contains two particles in a superposition of states $\ket{a}$ and $\ket{b}$. The following table presents the possible outcomes with corresponding probabilities for state $\ket{\psi}$:
\renewcommand{\arraystretch}{1.5}
\begin{table}[H]
\caption{Possible outcomes for state $\ket{\psi}$}
\centering
\begin{tabular}{ r | c | c | }
\multicolumn{1}{r}{}
& \multicolumn{1}{c}{Particle 2: $\ket{a}$}
& \multicolumn{1}{c}{Particle 2: $\ket{b}$} \\
\cline{2-3}
Particle 1: $\ket{a}$ & $\ket{\psi}\rightarrow\ket{aa}$: $\frac{1}{4}$ & $\ket{\psi}\rightarrow\ket{ab}$: $\frac{1}{4}$ \\
\cline{2-3}
Particle 1: $\ket{b}$ & $\ket{\psi}\rightarrow\ket{ba}$: $\frac{1}{4}$ & $\ket{\psi}\rightarrow\ket{bb}$: $\frac{1}{4}$ \\
\cline{2-3}
\end{tabular}
\end{table}
\noindent As an example measuring first particle yields state $\ket{a}$ with probability $\frac{1}{4}+\frac{1}{4}=0.5$. The next table illustrates the outcomes of measuring two particles in the second state $\ket{\phi}$:
\begin{table}[H]
\caption{Possible outcomes for state $\ket{\phi}$}
\centering
\begin{tabular}{ r | c | c | }
\multicolumn{1}{r}{}
& \multicolumn{1}{c}{Particle 2: $\ket{a}$}
& \multicolumn{1}{c}{Particle 2: $\ket{b}$} \\
\cline{2-3}
Particle 1: $\ket{a}$ & $\ket{\phi}\rightarrow\ket{aa}$: $\frac{1}{2}$ & $\ket{\phi}\rightarrow\ket{ab}$: 0 \\
\cline{2-3}
Particle 1: $\ket{b}$ & $\ket{\phi}\rightarrow\ket{ba}$: 0 & $\ket{\phi}\rightarrow\ket{bb}$: $\frac{1}{2}$ \\
\cline{2-3}
\end{tabular}
\end{table}
\noindent Measuring the first particle yields state $\ket{a}$ with probability 0.5, however now the measurement immediately collapsed the second particle state to the state $\ket{a}$. The individual states of two particles are related to each other and this relation is called quantum entanglement. The most astonishing part of quantum entanglement is that it does not depend on the distance. Assume we prepare two particles to be entangled and send the second one to other galaxy (without disturbing it). Then observer A measures the first particle on Earth and gets state $\ket{a}$. The observer B must also get state $\ket{a}$ measuring the second particle in other galaxy. The measurement outcome of the first particle will always immediately determine measurement outcome of the second particle. It does not mean that the information is moving faster than light, because to communicate results between observer A and observer B the classical communication channel still needs to be used, however once the results are compared we will always find out that the effects of quantum entanglement hold. The state is in quantum entanglement if it cannot be rewritten as a tensor product of single particle states. As an example, the state $\ket{\psi}$ could be rewritten as:
\begin{align}
\ket{\psi}&=\left(\frac{1}{\sqrt{2}}\ket{a}+\frac{1}{\sqrt{2}}\ket{b}\right)\otimes\left(\frac{1}{\sqrt{2}}\ket{a}+\frac{1}{\sqrt{2}}\ket{b}\right) \nonumber \\&=\frac{1}{2}\ket{aa}+\frac{1}{2}\ket{ba}+\frac{1}{2}\ket{ab}+\frac{1}{2}\ket{bb} \,.
\end{align}
For entangled state $\ket{\phi}$ this is not possible, meaning that the states are not separable. To prove this assume there exist complex numbers $c_1$, $c_2$ and $c_1^\prime$, $c_2^\prime$ so that:
\begin{align}
\ket{\phi}=\left(c_1\ket{a}+c_2\ket{b}\right)\otimes\left(c_1^\prime \ket{a}+c_2^\prime \ket{b}\right)=\frac{1}{\sqrt{2}}\ket{aa}+\frac{1}{\sqrt{2}}\ket{bb} \,.
\end{align}
However, this would imply that $c_1 c_1^\prime=c_2 c_2^\prime=\frac{1}{\sqrt{2}}$ and $c_1 c_2^\prime=c_2 c_1^\prime=0$, which obviously does not have any solution. Thus, the state $\ket{\phi}$ cannot be rewritten as a tensor product of single particle states.

\subsection{Density matrix}
There are some cases that states machinery is not a sufficient tool to describe a quantum system. In quantum mechanics we are considering two forms of probabilities :
\begin{itemize}
\item One that relates to the states and means probability of measuring an arbitrary state in one of eigenvectors of an observable $O$. The probability is derived from the norm square of probability amplitude $c_i$. This form is associated with pure states,
\item The other form relates to the whole system and means the probability of finding the particular system in one of the possible states. The probability expresses the fact that we do not have a full knowledge about the system and we can only say that the system is in $i$-th pure state with probability $p_i$. This form is associated with mixed states.
\end{itemize}
In order to describe both pure and mixed states a density matrix notation is used. A density matrix for pure state is defined as:
\begin{align}
\rho=\ket{\psi}\bra{\psi}
\end{align}
and for mixed state as:
\begin{align}
\rho=\sum_{i=1}^n p_i\ket{\psi_i}\bra{\psi_i}
\end{align}
where $p_i$ is interpreted as finding mixed state in the state $\ket{\psi_i}$. Actually, the density matrix for pure state is just a special case of density matrix for mixed state with $p_1=1$. The expression $\ket{\psi}\bra{\psi}$ is calculated as tensor product. To see how it is done we analyze the following pure state:
\begin{align}
\ket{\psi} = \frac{i}{2}\begin{bmatrix} 1 \\[2pt] 0 \end{bmatrix} + \frac{\sqrt{3}}{2}\begin{bmatrix} 0 \\[2pt] 1 \end{bmatrix} = \begin{bmatrix} \frac{i}{2} \\[2pt] \frac{\sqrt{3}}{2} \end{bmatrix}
\end{align}
then we take the tensor product to express density matrix:
\begin{align}
\rho&=\ket{\psi}\bra{\psi}=\begin{bmatrix} \frac{i}{2} \\[2pt] \frac{\sqrt{3}}{2} \end{bmatrix} \otimes \begin{bmatrix} \frac{-i}{2} & \frac{\sqrt{3}}{2} \end{bmatrix} \nonumber \\
&= \begin{bmatrix} \frac{i}{2} \frac{-i}{2} & \frac{i}{2}\frac{\sqrt{3}}{2} \\[2pt] \frac{\sqrt{3}}{2}\frac{-i}{2} & \frac{\sqrt{3}}{2}\frac{\sqrt{3}}{2}\end{bmatrix}= \begin{bmatrix} \frac{1}{4} & \frac{i\sqrt{3}}{4} \\[2pt] \frac{-i\sqrt{3}}{4} & \frac{3}{4}\end{bmatrix} \,.
\end{align}
From Eq.~(\ref{eq:expectationO}) in the previous section we know that the expectation value of state $\ket{\psi}$ given an observable $O$:
\begin{align}
O =\begin{bmatrix} 1 & 0 \\ 0 & 2 \end{bmatrix}
\end{align}
is equal to $1.75$. It turns out that the same result could be obtained using density matrix and following formula:
\begin{align}
\langle O \rangle &=\Tr{(\rho O}) \nonumber \\
&=\Tr{\left(\begin{bmatrix} \frac{1}{4} & \frac{i\sqrt{3}}{4} \\[2pt] \frac{-i\sqrt{3}}{4} & \frac{3}{4}\end{bmatrix}\begin{bmatrix} 1 & 0 \\[3pt] 0 & 2 \end{bmatrix}\right)}=\Tr{\left(\begin{bmatrix} \frac{1}{4} & \frac{2i\sqrt{3}}{4} \\[2pt] \frac{-i\sqrt{3}}{4} & \frac{6}{4}\end{bmatrix} \right)} \nonumber \\
&= \frac{1}{4} + \frac{6}{4}=1.75 \label{eq:trace}
\end{align}
yielding result as expected. As an example of mixed state, assume the situation in which we are unsure whether the system has been prepared in state $\ket{\psi}$ or $\ket{\phi}$, however we know the probability of system being in the first state is $p=\frac{1}{4}$ and being in the second state is $q=1-p=\frac{3}{4}$. The additional state $\ket{\phi}$ is defined as:
\begin{align}
\ket{\phi} = \frac{1}{\sqrt{2}}\begin{bmatrix} 1 \\[2pt] 0 \end{bmatrix} + \frac{1}{\sqrt{2}}\begin{bmatrix} 0 \\[2pt] 1 \end{bmatrix} = \begin{bmatrix} \frac{1}{\sqrt{2}} \\[2pt] \frac{1}{\sqrt{2}} \end{bmatrix} \,.
\end{align}
The expected value of this state given the same observable $O$ is $1.5$, as in half of the cases we will find the state in the first eigenvector with eigenvalue $\lambda_1=1$ and in half of the cases we will find the state in the second eigenvector with eigenvalue $\lambda_2=2$. It is not possible to represent a mixed state using state formalism, however it could be described with the density matrix:
\begin{align}
\rho&=p\ket{\psi}\bra{\psi}+q\ket{\phi}\bra{\phi}= \nonumber \\
&=\frac{1}{4}\begin{bmatrix} \frac{1}{4} & \frac{i\sqrt{3}}{4} \\[2pt] \frac{-i\sqrt{3}}{4} & \frac{3}{4}\end{bmatrix}+\frac{3}{4}\begin{bmatrix} \frac{1}{2} & \frac{1}{2} \\[3pt] \frac{1}{2} & \frac{1}{2}\end{bmatrix}= \begin{bmatrix} \frac{7}{16} & \frac{6+i\sqrt{3}}{16} \\[3pt] \frac{6-i\sqrt{3}}{16} & \frac{9}{16}\end{bmatrix} \,.
\end{align} 
For the mixed state we anticipate that in $25\%$ of outcomes the average will be $1.75$ and in $75\%$ of outcomes the average will be $1.5$, thus yielding expected value of $1.5625$. This result could be also obtained using exactly the same formula Eq.~(\ref{eq:trace}) as for pure states:
\begin{align}
\langle O \rangle &=\Tr{(\rho O}) \nonumber \\
&=\Tr{\left(\begin{bmatrix} \frac{7}{16} & \frac{6+i\sqrt{3}}{16} \\[3pt] \frac{6-i\sqrt{3}}{16} & \frac{9}{16}\end{bmatrix}\begin{bmatrix} 1 & 0 \\[3pt] 0 & 2 \end{bmatrix}\right)}= \frac{7}{16} + \frac{18}{16}=1.5625 \,.
\end{align}
Thus, the density matrix is a common tool for describing both pure and mixed states and allows us to use quantum mechanics in case we do not have full knowledge about the system. For further information about density matrices and other concepts in basic quantum mechanics see Ref.~\cite{Griffiths}.

\section{Quantum computation}

\subsection{Qubit}
The definition of qubit lies in the center of quantum computation theory. Before we dive into the explanation of qubit, let us recall the definition of classical bit. A bit is an unit of information, which describes a two-dimensional classical system. Thus, the classical system could be either in the state:
\begin{align}
\ket{0}=\begin{bmatrix} 1 \\ 0 \end{bmatrix}
\end{align}
or in the state:
\begin{align}
\ket{1}=\begin{bmatrix} 0 \\ 1 \end{bmatrix} \,.
\end{align}
The physical representation of a bit is two \textit{flip-flop} states representation, for instance two distinct voltages of electric circuit or two distict levels of light intensity. This is sufficient for classical physics and this how the classical computer works. The quantum computer uses the effects of quantum mechanics such as a superposition of states. A qubit is an unit of information, which describes a two-dimensional quantum system and the general state of qubit is represented by a pair of complex numbers:
\begin{align}
c_1\begin{bmatrix} 1 \\ 0 \end{bmatrix}+c_2\begin{bmatrix} 0 \\ 1 \end{bmatrix}=\begin{bmatrix} c_1 \\ c_2 \end{bmatrix}
\end{align}
so that it is a superposition of the states $\ket{0}$ and $\ket{1}$. The physical representation of qubit could be the polarization of a photon, spin of a particle or ground and exited orbit of an electron in atom. A classical computer handles with a string of bits for instance $01010001$ and based on it does the calculations. Quantum computation assumes that qubits can be assembled using tensor product, thus the same string could be written as:
\begin{align}
&\ket{0}\otimes\ket{1}\otimes\ket{0}\otimes\ket{1}\otimes\ket{0}\otimes\ket{0}\otimes\ket{0}\otimes\ket{1} \nonumber \\
&=\begin{bmatrix} 1 \\ 0 \end{bmatrix}\otimes\begin{bmatrix} 0 \\ 1 \end{bmatrix}\otimes\begin{bmatrix} 1 \\ 0 \end{bmatrix}\otimes\begin{bmatrix} 0 \\ 1 \end{bmatrix}\otimes\begin{bmatrix} 1 \\ 0 \end{bmatrix}\otimes\begin{bmatrix} 1 \\ 0 \end{bmatrix}\otimes\begin{bmatrix} 1 \\ 0 \end{bmatrix}\otimes\begin{bmatrix} 0 \\ 1 \end{bmatrix}
\end{align}
which is a vector with 256 rows:
\begin{align}
\ket{01010001}=\kbordermatrix{ & \\ 
\ket{00000000} & 0 \\ \ket{00000001} & 0 \\ \vdots & \vdots \\ \ket{01010000} & 0 \\ \ket{01010001} & 1 \\ \vdots & \vdots \\ \ket{11111110} & 0 \\ \ket{11111111} & 0
} \,.
\end{align}
Note that the general state of 8 qubit quantum computer could be written as a superposition:
\begin{align}
\ket{\psi} &= c_1\ket{00000000} + \dots + c_ {82}\ket{01010001} +\dots +c_{255} \ket{11111110} + c_{256}\ket{11111111} \nonumber \\
&= \sum_{i=1}^N c_i \ket{i}
\end{align}
where a set of $\ket{i}$ is called the standard computational basis and complex numbers are normalized, so that $\sum_{i=1}^N|c_i|^2=1$. This illustrates the overwhelming difference between quantum and classical computers. In order to write 8 qubit system 256 complex numbers are required, whereas on classical computer only 8 zeros or ones are required to fully describe 8 bit system. The effect increases exponentially with number of qubits, for 64 qubits we will need $2^{64}=18,446,744,073,709,551,616$ complex numbers to emulate quantum state on a classical machine. The quantum algorithms are exploiting this astonishing fact as we will see in the further sections. 

\subsection{Quantum gates}
\begin{figure}[h]
\begin{equation*}
\Qcircuit @C=1em @R=1em @!R {
\lstick{\ket{0}} & \gate{H} & \multigate{1}{\text{SWAP}} & \qw \\
\lstick{\ket{0}} & \qw & \ghost{\text{SWAP}} & \qw 
}
\end{equation*}
\caption{The operation $\ket{\psi_0} \rightarrow \ket{\psi_2}$ presented as quantum circuit.}
\label{fig:psi2}
\end{figure}
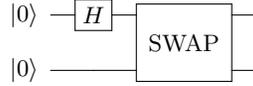

\begin{figure}[h]
\begin{equation*}
\Qcircuit @C=1em @R=1em @!R {
\lstick{\ket{0}} & \gate{H} & \multigate{1}{\text{SWAP}} & \meter \\
\lstick{\ket{0}} & \qw & \ghost{\text{SWAP}} & \meter 
}
\end{equation*}
\caption{The operation $\ket{\psi_0} \rightarrow \ket{\psi_2}$ and measurement presented as quantum circuit.}
\label{fig:psi2measurement}
\end{figure}
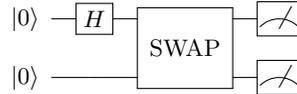

A classical logical gate is a way of bits manipulation. As an example the gate \textit{NOT} flips the bit so that $\textit{NOT}\ket{0}= \ket{1}$ and $\textit{NOT}\ket{1} = \ket{0}$. This can be represented by following 2-by-2 matrix:
\begin{align}
\textit{NOT}=\begin{bmatrix} 0 & 1 \\ 1 & 0 \end{bmatrix} \,. \label{eq:not}
\end{align}
The matrix defined in Eq.~(\ref{eq:not}) satisfies:
\begin{align}
\begin{bmatrix} 0 & 1 \\ 1 & 0 \end{bmatrix} \begin{bmatrix} 1 \\ 0 \end{bmatrix} = \begin{bmatrix} 0 \\ 1 \end{bmatrix}
\end{align}
and
\begin{align}
\begin{bmatrix} 0 & 1 \\ 1 & 0 \end{bmatrix} \begin{bmatrix} 0 \\ 1 \end{bmatrix} = \begin{bmatrix} 1 \\ 0 \end{bmatrix} \,.
\end{align}
The second example of a classical gate is \textit{AND} gate, which accepts two bits and outputs one. The \textit{AND} gate is represented by 2-by-4 matrix:
\begin{align}
\textit{AND}=\begin{bmatrix} 1 & 1 & 1 & 0 \\ 0 & 0 & 0 & 1 \end{bmatrix}
\end{align}
and satisfies the following relations:
\begin{itemize}
\item $\textit{AND}\ket{00}=\ket{0}$,
\item $\textit{AND}\ket{01}=\ket{0}$,
\item $\textit{AND}\ket{10}=\ket{0}$,
\item $\textit{AND}\ket{11}=\ket{1}$.
\end{itemize}
Quantum gates are a way of qubits manipulation. A state enters a gate in quantum circuit and exits as other state, thus quantum gates represent time evolution of a state describing qubits. The quantum gate satisfies the following criteria:
\begin{itemize}
\item must preserve norms i.e. norm squared probability amplitudes sum to one after gate application,
\item must be reversible i.e. evolution of each not measured quantum state must be reversible.
\end{itemize}
These conditions are equivalent to a restriction that quantum gates must be unitary matrices. Thus, the 2x4 \text{AND} gate is not a valid quantum gate, because it is not unitary, whereas \text{NOT} gate is a valid quantum gate also known as one of the three Pauli matrices used in quantum mechanics:
\begin{align}
\sigma_1=\begin{bmatrix} 0 & 1 \\ 1 & 0 \end{bmatrix}, \phantom{000} \sigma_2=\begin{bmatrix} 0 & -i \\ i & 0 \end{bmatrix}, \phantom{000} \sigma_3=\begin{bmatrix} 1 & 0 \\ 0 & -1 \end{bmatrix} \,.
\end{align}
Another trivial example of unitary matrix is the identity matrix:
\begin{align}
I=\begin{bmatrix} 1 & 0 \\ 0 & 1 \end{bmatrix} \,.
\end{align}
Frequently used gate in quantum computation is Hadamard gate which allows to produce superposition of states:
\begin{align}
H=\frac{1}{\sqrt{2}}\begin{bmatrix} 1 & 1 \\ 1 & -1 \end{bmatrix} \,.
\end{align}
Applying Hadamard gate on qubit in state $\ket{0}$ yields:
\begin{align}
H\ket{0}=\frac{1}{\sqrt{2}}\begin{bmatrix} 1 & 1 \\ 1 & -1 \end{bmatrix}\begin{bmatrix} 1 \\ 0 \end{bmatrix}=\frac{1}{\sqrt{2}}\begin{bmatrix} 1 \\ 1 \end{bmatrix}=\frac{\ket{0}+\ket{1}}{\sqrt{2}}
\end{align}
and on qubit in state $\ket{1}$:
\begin{align}
H\ket{1}=\frac{1}{\sqrt{2}}\begin{bmatrix} 1 & 1 \\ 1 & -1 \end{bmatrix}\begin{bmatrix} 0 \\ 1 \end{bmatrix}=\frac{1}{\sqrt{2}}\begin{bmatrix} 1 \\ -1 \end{bmatrix}=\frac{\ket{0}-\ket{1}}{\sqrt{2}} \,.
\end{align}
There are also quantum gates acting on two qubits, for instance \textit{SWAP} gate represented by:
\begin{align}
\textit{SWAP}=\begin{bmatrix} 1 & 0 & 0 & 0 \\ 0 & 0 & 1 & 0 \\ 0 & 1& 0 & 0 \\ 0 & 0 & 0 & 1 \end{bmatrix} \,.
\end{align}
It swaps two qubits so that the example state $\ket{01}$ is evolved into state $\ket{10}$:
\begin{align}
\textit{SWAP}\ket{01}=\begin{bmatrix} 1 & 0 & 0 & 0 \\ 0 & 0 & 1 & 0 \\ 0 & 1& 0 & 0 \\ 0 & 0 & 0 & 1 \end{bmatrix}\begin{bmatrix} 0 \\ 1 \\ 0 \\ 0 \end{bmatrix} = \begin{bmatrix} 0 \\ 0 \\ 1\\ 0 \end{bmatrix} = \ket{10} \,.
\end{align}
We can apply quantum gates on more than one qubit using tensor product. As an example we initialize two qubits in state $\ket{\psi_0}=\ket{00}$, apply Hadamard gate only on the first qubit leaving the second qubit unchanged and resulting in state $\ket{\psi_1}$:
\begin{align}
\ket{\psi_1}&=(H \otimes I)\ket{00} \nonumber \\
&=\frac{1}{\sqrt{2}}\left(\begin{bmatrix} 1 & 1 \\ 1 & -1 \end{bmatrix} \otimes \begin{bmatrix} 1 & 0 \\ 0 & 1 \end{bmatrix}\right) \left(\begin{bmatrix} 1 \\ 0 \end{bmatrix}\otimes\begin{bmatrix} 1 \\ 0 \end{bmatrix}\right) \nonumber \\
&=\frac{1}{\sqrt{2}}\begin{bmatrix} 1 & 0 & 1 & 0\\0 & 1 & 0 & 1\\1 & 0 & -1 & 0\\0 & 1 & 0 & -1 \end{bmatrix} \begin{bmatrix} 1 \\ 0 \\ 0 \\ 0 \end{bmatrix}=\frac{1}{\sqrt{2}}\begin{bmatrix} 1 \\ 0 \\ 1 \\ 0 \end{bmatrix} \nonumber \\
&=\frac{1}{\sqrt{2}}(\ket{00}+\ket{10})=\frac{\ket{0}+\ket{1}}{\sqrt{2}} \otimes \ket{0} \,.
\end{align}
In general we can combine any number of quantum gates using tensor products and use it on qubit state as long as the dimensions of the state vector matches the size of combined quantum gate. The operation $\ket{\psi_0} \rightarrow \ket{\psi_1}$ can be presented in quantum circuit:
\begin{figure}[h]
\begin{equation*}
\Qcircuit @C=1em @R=1em @!R {
\lstick{\ket{0}} & \gate{H} & \qw \\
\lstick{\ket{0}} & \qw & \qw 
}
\end{equation*}
\caption{The operation $\ket{\psi_0} \rightarrow \ket{\psi_1}$ presented as quantum circuit.}
\end{figure}
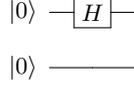
\\
Continuing the example, the state $\ket{\psi_1}$ could be evolved to $\ket{\psi_2}$ using \textit{SWAP} gate on two qubits in a superposition:
\begin{align}
\ket{\psi_2}&=\textit{SWAP}\left(\frac{\ket{0}+\ket{1}}{\sqrt{2}} \otimes \ket{0}\right) \nonumber \\
&=\frac{1}{\sqrt{2}}\begin{bmatrix} 1 & 0 & 0 & 0 \\ 0 & 0 & 1 & 0 \\ 0 & 1& 0 & 0 \\ 0 & 0 & 0 & 1 \end{bmatrix}\begin{bmatrix} 1 \\ 0 \\ 1 \\ 0 \end{bmatrix}=\frac{1}{\sqrt{2}}\begin{bmatrix} 1 \\ 1 \\ 0 \\ 0 \end{bmatrix} \nonumber \\
&=\frac{1}{\sqrt{2}}(\ket{00}+\ket{01})=\ket{0} \otimes \frac{\ket{0}+\ket{1}}{\sqrt{2}}
\end{align}
As expected the \textit{SWAP} gate interchanged the qubits. This operation can be presented in quantum circuit in Fig.~(\ref{fig:psi2}).
\\
The last step is to measure both qubits. Measurement is an irreversible operation, thus is not a quantum gate, but a final step of quantum circuit. Measuring the first qubit we will always yield state $\ket{0}$ and measuring the second qubit we have $50\%$ chances of getting qubit in state $\ket{0}$ and $50\%$ chances of getting qubit in state $\ket{1}$. The measurement can be illustrated in Fig.~(\ref{fig:psi2measurement}) using meter symbol.
\\
The quantum computation can be defined as a unitary evolution of a closed quantum system - it takes some initial state as an input and outputs a final state, which can be measured to retrieve specific information. It is crucial that a quantum system remains isolated, i.e. there is no deconherence. Any deconherence is a loss of information from a system into the environment and relevant quantum benefits can vanish. Interestingly, the quantum computation using small number of qubits can be simulated on classical computer with a few software solutions being open-source \cite{Quirk,QuTiP}.

\subsection{Quantum parallelism}
\begin{figure}[h]
\begin{equation*}
\Qcircuit @C=1em @R=1em @!R {
\lstick{\ket{\phi}} & \multigate{1}{\text{O}} & \qw \\
\lstick{\ket{0}} & \ghost{\text{O}} & \qw
}
\end{equation*}
\caption{Application of quantum gate \textit{O} presented as quantum circuit.}
\label{circuitPar}
\end{figure}
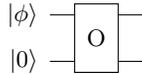

The quantum speed-up of algorithm comes from quantum parallelism achieved through the superposition of qubits. The quantum parallelism is a feature of many quantum algorithms and in simple words allows to evaluate function on many inputs simultaneously. Suppose we have a function $f(\phi): \{0,1\}^n \rightarrow \{0,1\}$. We initialize a quantum system in the state $\ket{\psi_0}=\ket{\phi,0}$ and evolve it to the state $\ket{\psi_1}=O\ket{\phi,0}$ with an appropriate sequence of quantum gates, which combined can be represented by gate \textit{O}. The quantum gate \textit{O} takes the input $\ket{x,0}$ and evolves it to $\ket{x,f(x)}$. For simplicity, assume that $n=2$ and our first register with state $\ket{\phi}$ is initialized as:
\begin{align}
\ket{\phi}=\frac{\ket{0}+\ket{1}}{\sqrt{2}} \otimes \frac{\ket{0}+\ket{1}}{\sqrt{2}}
\end{align}
which can be achieved by application of Hadamard gate on both qubits $\ket{00}$ (however this operation is assumed to be done before in a separate quantum circuit). The quantum circuit is presented in Fig.~\ref{circuitPar}.
\vspace*{-1em}
\\
To summarize we started with the state $\ket{\psi_0}$:
\begin{align}
\ket{\psi_0}=\ket{\phi,0}=\frac{1}{2}\left(\ket{00}+\ket{01}+\ket{10}+\ket{11}\right) \otimes \ket{0}
\end{align}
and ended with the state $\ket{\psi_1}$:
\begin{align}
\ket{\psi_1}&=O\ket{\phi,0}=\frac{1}{2}\left(O\ket{00,0}+O\ket{01,0}+O\ket{10,0}+O\ket{11,0}\right) \nonumber \\ 
&=\frac{1}{2}\left(\ket{00,f(00)}+\ket{01,f(01)}+\ket{10,f(10)}+\ket{11,f(11)}\right) \,.
\end{align}
Thus, with just one operation we evaluated function $f$ for four inputs $00$, $01$, $10$ and $11$ simultaneously. The same operation on a classical computer must be executed one by one for each input or done in separate classical circuits. This remarkable feature of quantum algorithms uses the fact that input state is in superposition. The general logic can be expanded for large $n$ making algorithm more effective. The only disadvantage is that the outputs of function are stored as quantum states, thus measuring the second qubit yields only one evaluated value. Nevertheless, the quantum parallelism is effectively used by many quantum algorithms as an intermediate step in quantum circuit. 

\section{Quantum algorithms}
\begin{figure}[H]
\centering
\includegraphics[width=0.4\textwidth]{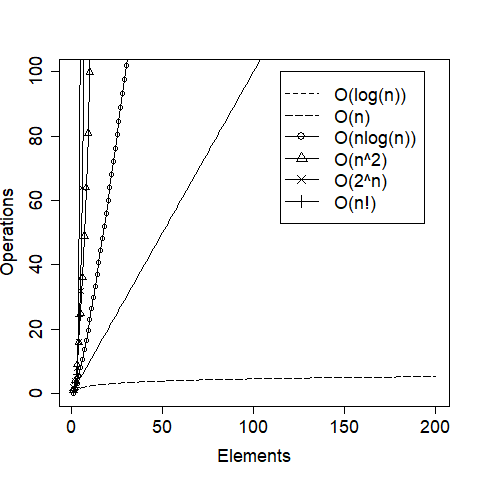}
\caption{Big-O notation}
\label{fig:bigO}
\end{figure}

This section presents quantum algorithms, which will be used by quantum machine learning algorithms as subroutines. The quantum algorithms presented are:
\begin{itemize}
\item Grover's search algorithm (basis for quantum minimization algorithm),
\item Quantum minimization algorithm,
\item Quantum Fourier transform algorithm (basis for quantum phase estimation algorithm),
\item Quantum phase estimation algorithm.
\end{itemize}
In theory, each quantum algorithm is either quadratically or exponentially faster than its classical counterpart. To measure algorithm efficiency the time complexities are used along with big-O notation. The big-O notation explains how fast time complexity of an algorithm grows when a main element is arbitrarily large. The main element is meant to be a size of input, a number of quantum gates used or a number of iterations used depending on which factor impacts the effectiveness of algorithm most. It presents the limiting behavior of an algorithm avoiding constants and not significant terms, which are much smaller in comparison to the main element that drives the algorithm time complexity. The Fig.~\ref{fig:bigO} gives the intuition about the big-O notation and indicates the order of time complexities.
Each algorithm is described in the following manner: a brief introduction in preliminaries, a step by step algorithm explanation  based on a simple example, then a general description of quantum algorithm and finally the conclusions about time complexity and other remarks in summary. 

\subsection{Grover's search algorithm}

\subsubsection{Preliminaries}
Grover's algorithm is a quantum search algorithm, which runtime is quadratically faster than any classical counterpart \cite{Grover}. Grover's algorithm is not intended to find an element in a database, its purpose is searching through function inputs to check whether the function returns true for that input. It is very efficient in case the function is unknown or extremely complex and we want to know for which input the function returns true or for which input the equation is solved. This function can be represented as a quantum oracle and may be built from a large number of combined quantum gates. It is assumed that the quantum oracle has been already given to the algorithm and serves as a black-box. As an example, imagine that we are given a function $f(x): \{0,1\}^n \rightarrow \{0,1\}$ as a set of \textit{AND} and \textit{OR} logical operators which for exactly one binary string of zeros and ones returns true. Coding such a function can be relatively simple, however examining for which combination of zeros and ones the function returns true would require at worst scenario $2^n$ function calls on a classical machine. On a quantum computer, we can transform the function into a valid set of quantum gates constituting for a quantum oracle $O$ and use Grover's search algorithm to find a correct input with very high accuracy in only $\sqrt{2^n}$ iterations.

\subsubsection{Example}
Grover's search algorithm relies on setting a state of the system into a superposition of all possible inputs and then, the probability of finding the searched input is increased in each algorithm iteration. We aim to explain the algorithm starting with a simple $n=2$ dimensional example. As an input, the algorithm requires a quantum oracle $O$ that \textit{picks out} the string $10$. The function representing that operation returns true if input is equal to searched binary string:
\begin{align}
f(x)=
\begin{cases}
1 & x=10\\
0 & x\neq 10
\end{cases} \,.
\end{align}
Then we need to represent function $f(x): \{0,1\}^2 \rightarrow \{0,1\}$ as quantum oracle $O$. This is a little bit tricky as first guess $O\ket{x}=\ket{f(x)}$ is not unitary and reversible operation, thus is not a quantum gate (the state $\ket{x}$ consists of $2$ qubits, whereas $\ket{f(x)}$ of only one qubit). However, there exists a method of constructing the quantum oracle as:
\begin{align}
O\ket{x}=(-1)^{f(x)}\ket{x} \,. \label{eq:oracle}
\end{align}
The flip of amplitude sign in case $f(x)=1$ \textit{marks} the state 10 and the matrix representing this operation is unitary, so that $O$ is desired quantum oracle. In that particular example the quantum oracle acts on two input qubits and is expressed as matrix:
\begin{align}
O = \kbordermatrix{ & \ket{00} & \ket{01} & \ket{10} & \ket{11} \\ 
\ket{00} & 1 & 0 & 0 & 0 \\ \ket{01} & 0 & 1 & 0 & 0 \\ \ket{10} & 0 & 0 & -1 & 0 \\ \ket{11} & 0 & 0 & 0 & 1
} \,.
\end{align}
At this point it is important to realize that building quantum oracle for that example has already unveiled for which input the function returns true. However, this is just for presentation purposes and in practice we have a black-box quantum oracle that is already given to the algorithm. \par
Obviously, the algorithm searches for the string $x=10$, thus we require that the measurement of the system at the end should reveal the correct answer with high probability. The detailed steps of the algorithm for that particular example are explained below. \\\\
\textbf{Initialize} \\
Initialize $\ket{0}^{\otimes2}$ qubits resulting in the state $\ket{\psi_0}=\ket{00}$. \\\\
\textbf{Apply Hadamard gates} \\
Apply Hadamard gate on each qubit resulting in the state:
\begin{align}
\ket{\psi_1}&=H^{\otimes2}\ket{\psi_0}=\frac{1}{\sqrt{2^2}}(\ket{0}+\ket{1})\otimes(\ket{0}+\ket{1}) \nonumber \\&=\frac{1}{\sqrt{2^2}}(\ket{00}+\ket{01}+\ket{10}+\ket{11})=\begin{bmatrix} \frac{1}{2} & \frac{1}{2} &\frac{1}{2} & \frac{1}{2} \end{bmatrix}^T \,.
\end{align}
After multiplication it represents all of the possible inputs to quantum oracle in equally weighted superposition. We used only 2 qubits to represent 4 inputs. \\\\
\textbf{Iteration (1): 1. Apply quantum oracle $O$ gate} \\
Amplitude sign of the searched input is flipped with this operation resulting in the state:
\begin{align}
\ket{\psi_2^{(1)}}=O\ket{\psi_1}=\frac{1}{\sqrt{2^2}}(\ket{00}+\ket{01}-\ket{10}+\ket{11})=\begin{bmatrix} \frac{1}{2} & \frac{1}{2} &-\frac{1}{2} & \frac{1}{2} \end{bmatrix}^T \,.
\end{align}
The quantum oracle is not enough to recognize the searched input, because the sign of amplitude has no effect on measurement probability. We have to look for additional quantum gates that makes the absolute value of amplitude greater for the searched state. \\\\
\textbf{Iteration (1): 1. Inversion around the mean} \\
In order to strengthen amplitude with negative sign we use a procedure called inversion around the mean. The operation is fairly simple and implies using $2A-I^{\otimes 2}$ gate, where $A$ is defined as:
\begin{align}
A= \begin{bmatrix} \frac{1}{2} & \frac{1}{2} & \frac{1}{2} & \frac{1}{2} \\[6pt] \frac{1}{2} & \frac{1}{2} & \frac{1}{2} & \frac{1}{2} \\[6pt]
\frac{1}{2} & \frac{1}{2} & \frac{1}{2} & \frac{1}{2} \\[6pt] \frac{1}{2} & \frac{1}{2} & \frac{1}{2} & \frac{1}{2} \end{bmatrix}
\end{align}
and the inversion around mean is represented by:
\begin{align}
2A-I^{\otimes 2}= \begin{bmatrix} -\frac{1}{2} & \frac{1}{2} & \frac{1}{2} & \frac{1}{2} \\[6pt] \frac{1}{2} & -\frac{1}{2} & \frac{1}{2} & \frac{1}{2} \\[6pt]
\frac{1}{2} & \frac{1}{2} & -\frac{1}{2} & \frac{1}{2} \\[6pt] \frac{1}{2} & \frac{1}{2} & \frac{1}{2} & -\frac{1}{2} \end{bmatrix} \,.
\end{align}
The gate has following action on the $\ket{\psi_2}$:
\begin{align}
\ket{\psi_3^{(1)}}=(2A-I^{\otimes 2})\ket{\psi_2^{(1)}} = \begin{bmatrix} -\frac{1}{2} & \frac{1}{2} & \frac{1}{2} & \frac{1}{2} \\[6pt] \frac{1}{2} & -\frac{1}{2} & \frac{1}{2} & \frac{1}{2} \\[6pt]
\frac{1}{2} & \frac{1}{2} & -\frac{1}{2} & \frac{1}{2} \\[6pt] \frac{1}{2} & \frac{1}{2} & \frac{1}{2} & -\frac{1}{2} \end{bmatrix} \begin{bmatrix} \frac{1}{2} \\[6pt] \frac{1}{2} \\[6pt] -\frac{1}{2} \\[6pt] \frac{1}{2} \end{bmatrix}=\begin{bmatrix} 0 \\[6pt] 0 \\[6pt] 1 \\[6pt] 0 \end{bmatrix} \,.
\end{align}
The searched input has probability amplitude of one, however this is not a general case for number of input qubits greater than two. In that cases the steps: application of quantum oracle gate and application of inversion around mean are repeated around $\frac{\pi}{4}\sqrt{2^n}$ times \cite{Grover}. As a quick example you can check that for $3$-qubit inputs (with the second element being searched) the two iterations in the algorithm would follow:
\begin{align}
& \ket{\psi_2^{(1)}} = \begin{bmatrix} \frac{\sqrt{2}}{4} & - \frac{\sqrt{2}}{4} & \frac{\sqrt{2}}{4} & \frac{\sqrt{2}}{4} & \frac{\sqrt{2}}{4} & \frac{\sqrt{2}}{4} & \frac{\sqrt{2}}{4} & \frac{\sqrt{2}}{4} \end{bmatrix}^T \nonumber \\
&\rightarrow\ket{\psi_3^{(1)}} = \begin{bmatrix} \frac{\sqrt{2}}{8} & \frac{5\sqrt{2}}{8} & \frac{\sqrt{2}}{8} & \frac{\sqrt{2}}{8} & \frac{\sqrt{2}}{8} & \frac{\sqrt{2}}{8} & \frac{\sqrt{2}}{8} & \frac{\sqrt{2}}{8} \end{bmatrix}^T \nonumber \\
&\rightarrow\ket{\psi_2^{(2)}} = \begin{bmatrix} \frac{\sqrt{2}}{8} & - \frac{5\sqrt{2}}{8} & \frac{\sqrt{2}}{8} & \frac{\sqrt{2}}{8} & \frac{\sqrt{2}}{8} & \frac{\sqrt{2}}{8} & \frac{\sqrt{2}}{8} & \frac{\sqrt{2}}{8} \end{bmatrix}^T \nonumber \\
&\rightarrow \ket{\psi_3^{(2)}} = \begin{bmatrix} \frac{\sqrt{2}}{16} & - \frac{11\sqrt{2}}{16} & \frac{\sqrt{2}}{16} & \frac{\sqrt{2}}{16} & \frac{\sqrt{2}}{16} & \frac{\sqrt{2}}{16} & \frac{\sqrt{2}}{16} & \frac{\sqrt{2}}{16} \end{bmatrix}^T 
\end{align}
so that the absolute value of amplitude for the second element is clearly higher. The equivalent representation of inversion around mean gate that is implemented in quantum circuit implies using gates:
\begin{align}
2A-I^{\otimes 2}=H^{\otimes 2}(2\ket{0^{\otimes2}}\bra{0^{\otimes2}}-I^{\otimes2})H^{\otimes 2} \,.
\end{align}
\textbf{Measurement} \\
The last step implies collapse of the state:
\begin{align}
\ket{\psi_3}=\begin{bmatrix} 0 & 0 & 1 & 0 \end{bmatrix}^T=0\ket{00}+0\ket{01}+1\ket{10}+0\ket{11}
\end{align}
by making measurement on both qubits. In that particular example the square of amplitude equals one, so that the searched input is always measured.

\newpage
\subsubsection{Algorithm}
Generally, the algorithm could be visualized by the quantum circuit presented in Fig.~(\ref{fig:circuitgrover}). Grover search algorithm requires a quantum oracle $O$ acting on $n$ input qubits. Measurement of the system at the end of the algorithm should reveal the searched string $x$ with high probability.
\begin{figure}[h] 
\begin{equation*}
\Qcircuit @C=1em @R=0.3em {
& & & & & & & \ustick{\text{Inversion around mean}} & & & & & \\
\lstick{\ket{0}} & /^n \qw & \gate{H^{\otimes n}} & \qw & \gate{O} & \qw & \gate{H^{\otimes n}} & \gate{2\ket{0^{\otimes2}}\bra{0^{\otimes2}}-I^{\otimes2}} & \gate{H^{\otimes n}} & \qw & \cdots & & \meter \\
& & & & & & & \dstick{\text{Repeat $\approx \frac{\pi}{4}\sqrt{2^n}$ times}} & & & & & \text{}
\gategroup{2}{7}{2}{9}{.7em}{^\}} \gategroup{2}{5}{2}{9}{.7em}{_\}}
}
\end{equation*} 
\caption{Quantum circuit of Grover's algorithm.}
\label{fig:circuitgrover}
\end{figure}
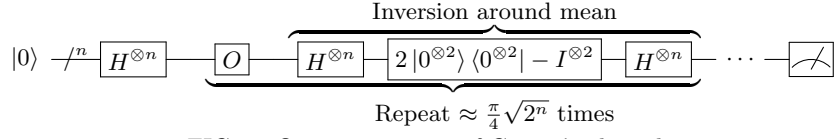 \\
\textbf{Initialize} \\
Initialize $\ket{0}^{\otimes n}$ qubits resulting in the state $\ket{\psi_0}=\ket{0\dots 0}$. \\\\
\textbf{Apply Hadamard gates} \\
Apply Hadamard gate on each qubit resulting in the state:
\begin{align}
\ket{\psi_1}=H^{\otimes n}\ket{\psi_0}=\frac{1}{\sqrt{2^n}}(\ket{0\dots00}+\ket{0\dots01}+\dots+\ket{1\dots11}) \,.
\end{align} \\
\textbf{Iteration (i): 1. Apply quantum oracle $O$ gate} \\
The amplitude sign of searched input is flipped with this operation resulting in the state:
\begin{align}
\ket{\psi_2^{(i)}}=O\ket{\psi_1} \,.
\end{align} \\
\textbf{Iteration (i): 2. Inversion around the mean} \\
Apply inversion around the mean gate:
\begin{align}
\ket{\psi_3^{(i)}}=(H^{\otimes n}(2\ket{0^{\otimes n}}\bra{0^{\otimes n}}-I^{\otimes n})H^{\otimes n})\ket{\psi_2} \,.
\end{align}
Repeat steps: Apply quantum oracle $O$ gate and inversion around the mean about $i_{max}\approx \frac{\pi}{4}\sqrt{2^n}$ times. \\\\
\textbf{Measurement} \\
Measure the state $\ket{\psi_3^{(i_{max})}}$. The searched item should be measured with a high probability. 

\subsubsection{Summary}
The quantum parallelism of Grover's search algorithm relies on changing amplitudes of all inputs simultaneously. This is done thanks to the superposition of states, which is purely a quantum concept. Moreover, the search is done globally, which indicates a significant improvement in optimization routines and in fact this is explored in the next section. It is known that the quantum algorithm is optimal, which means that quantum computers cannot do better in these type of problems \cite{Bennett}. Grover's search algorithm time complexity is around $O(\sqrt{2^n})$. In the case of a classical computer, this would be $O(2^n)$, which means a quadratic speed-up of the searching algorithm. It is not yet a holy-grail exponential speed-up, however it is still significant for large data vectors. On the other hand, Grover's algorithm is sensitive to the number of iterations. More iterations will decrease the amplitude of correct answer, thus incorrectly selecting this parameter can {\it overcook} the solution. Moreover, the operation of algorithm is limited in case of introducing a noise to the quantum system which is a reality in today's quantum computers \cite{Coles}. There also exists extension of Grover's search algorithm for finding $k$ matching entries instead of one \cite{Grover}. 

\subsection{Quantum minimization algorithm} \label{qGroverOptim}

\subsubsection{Preliminaries}
Grover's search algorithm is a special case of the optimization problem, as it finds an input for which $f(x)=1$ or in other words global maximum of a function returning zeros or one. This is extended in the quantum minimization algorithm, known also as D{\"u}rr and H{\o}yer minimization algorithm to deal with more general problems \cite{Durr}. The quantum algorithm uses Grover's search algorithm (version for $k$ matching entries) with quantum oracles indicating which elements are smaller than an arbitrary threshold. Grover's search is performed several times in order to find a solution to the optimization problem. More importantly, thanks to a quantum parallelism global minimum is always found, which has large implications for machine learning procedures, as they often stuck in local optima. 

\subsubsection{Example}
\noindent We would minimize an objective function $f(x): \{0,1\}^3 \rightarrow \{0,1,2,3\}$:
\begin{align}
f(x)=
\begin{cases}
0 & x=100\\
1 & x=000\\
2 & x=001\\
3 & otherwise
\end{cases} \,,
\end{align}
As an input a quantum oracle $O$ is required that \textit{marks} inputs for which function is lower than a threshold $y$. For simplicity we assume that $y=\{0,1,2,3\}$. The function representing that operation returns true if $f(x)$ is lower than $y$:
\begin{align}
h(x)=
\begin{cases}
1 & f(x)<y\\
0 & f(x) \geq y 
\end{cases} \,.
\end{align}
Then we use a standard prescription defined in Eq.~(\ref{eq:oracle}) to build a quantum oracle:
\begin{align}
O\ket{x}=(-1)^{h(x)}\ket{x} \,.
\end{align}
This time Grover's search algorithm finds $k$ matching entries, so that several amplitude signs may be changed. The simplest matrix representation of quantum oracle can be written as:
\begin{align}
O=\delta(y-0)O_0+\delta(y-1)O_1+\delta(y-2)O_2+\delta(y-3)O_3
\end{align}
where $\delta(x-a)$ is Dirac delta function returning one if $x=a$ and zero otherwise. The component $O_2$ is written as:
\begin{align}
O_2 &=\kbordermatrix{ & \ket{000} & \ket{001} & \ket{010} & \ket{011} & \ket{100} & \ket{101} & \ket{110} & \ket{111} \\ 
\ket{000} & -1 & 0 & 0 & 0 & 0 & 0 & 0 & 0 \\ \ket{001} & 0 & 1 & 0 & 0 & 0 & 0 & 0 & 0 \\ \ket{010} & 0 & 0 & 1 & 0 & 0 & 0 & 0 & 0 \\ \ket{011} & 0 & 0 & 0 & 1 & 0 & 0 & 0 & 0 \\
\ket{100} & 0 & 0 & 0 & 0 & -1 & 0 & 0 & 0 \\ \ket{101} & 0 & 0 & 0 & 0 & 0 & 1 & 0 & 0 \\ \ket{110} & 0 & 0 & 0 & 0 & 0 & 0 & 1 & 0 \\ \ket{111} & 0 & 0 & 0 & 0 & 0 & 0 & 0 & 1 
} 
\end{align}
so that it flips the amplitude sign for inputs where $f(x)<2$ i.e. $x=100$ and $x=000$. Other components $O_0$, $O_1$ and $O_3$ are represented analogously. Similarly to Grover's search algorithm it is important to realize that building quantum oracle for that example already unveils for which input the function has minimum. However, this is just for presentation purposes and in practice we have a black-box quantum oracle that is already given to the algorithm. \par
The correct answer $x_{min}=100$ for which function has global minimum should be measured with high probability. The detailed steps of the algorithm for that particular example are explained below. 

\newpage
\noindent \textbf{Set starting point} \\
Randomly select input $x_1$ and set $y_1=f(x_1)$ as a starting point of the algorithm. In our example we randomly selected $x_1=001$ for which $y_1=2$. By expressing function outputs as binary strings:
\begin{align}
0 &\rightarrow 00 \nonumber \\
1 &\rightarrow 01 \nonumber \\
2 &\rightarrow 10 \nonumber \\
3 &\rightarrow 11 
\end{align}
and then converting them to quantum states, a digit 2 can be represented by the state $\ket{10}$. \\\\
\textbf{Iteration (1): 1. Initialize} \\
Initialize the first register with $3$ input qubits and the second register with $2$ qubits to store threshold $y_1$. The initial state can be written as:
\begin{align}
\ket{\psi_0^{(1)}}=\ket{000}\ket{10} \,.
\end{align}
\textbf{Iteration (1): 2. Apply Hadamard gates} \\
Apply Hadamard gates on the first register resulting in the state:
\begin{align}
\ket{\psi_1^{(1)}}&=(H^{\otimes 3}\otimes I^{\otimes 2})\ket{\psi_1}=\frac{1}{\sqrt{2^3}}(\ket{000}+\ket{001}+\dots+\ket{111})\ket{10} \nonumber \\ 
&= \begin{bmatrix} \frac{\sqrt{2}}{2} & \frac{\sqrt{2}}{2} & \frac{\sqrt{2}}{2} & \frac{\sqrt{2}}{2} & \frac{\sqrt{2}}{2} & \frac{\sqrt{2}}{2} & \frac{\sqrt{2}}{2} & \frac{\sqrt{2}}{2} \end{bmatrix}^T \otimes \begin{bmatrix} 0 \\ 0 \\ 1 \\ 0 \end{bmatrix} \,.
\end{align} 
The first register represents all possible inputs to function $f(x)$. \\\\
\textbf{Iteration (1): 3. Apply Grover's search algorithm} \\
Apply Grover's algorithm using quantum oracle $O$ and inversion around mean $r$ times. The value of $r$ is chosen randomly in order to avoid algorithm trapping \cite{Baritompa}. In our example Grover's algorithm (with $r_1=1$) produces the state:
\begin{align}
\ket{\psi_2^{(1)}}&=(G \otimes I^{\otimes 2})\frac{1}{\sqrt{2^3}}(\ket{000}+\ket{001}+\dots+\ket{111})\ket{10} \nonumber \\
&= \begin{bmatrix} \frac{1}{\sqrt{2}} & 0 & 0 & 0 & \frac{1}{\sqrt{2}} & 0 & 0 & 0 \end{bmatrix}^T \otimes \begin{bmatrix} 0 \\ 0 \\ 1 \\ 0 \end{bmatrix} \,.
\end{align}
The state after measurement of the first register will either collapse to $\ket{100}\ket{10}$ or $\ket{000}\ket{10}$ with 0.5 probability for both outcomes. In our example assume that the measurement resulted in the state $\ket{000}\ket{10}$. \\\\
\textbf{Iteration (1): 4. Set new threshold} \\
Set $x_2=000$ and $y_2=f(000)=1$. Change the second register to $\ket{01}$ corresponding to a digit 1. Go to the next iteration. \\\\
\textbf{Convergence} \\
The convergence for that example is achieved after the second iteration. In the second iteration Grover's algorithm (with $r_2=2$) produces the state (amplitudes rounded to third digit):
\begin{align}
\ket{\psi_2^{(2)}}= \begin{bmatrix} 0.972 & -0.088 & -0.088 & -0.088 & -0.088 & -0.088 & -0.088 & -0.088 \end{bmatrix}^T \otimes \begin{bmatrix} 0 \\ 1 \\ 0 \\ 0 \end{bmatrix}
\end{align}
and the global minimum $x=100$ is measured with probability of $0.972^2 = 0.945$. Advice on a number of iterations is presented in the general algorithm description. 

\subsubsection{Algorithm}
\begin{figure}[h]
\begin{equation*}
\Qcircuit @C=1em @R=1em {
\lstick{\ket{0}} & /^n \qw & \gate{H^{\otimes n}} & \qw & \gate{G_{r_i}} & \qw & \meter \\
\lstick{\ket{y_i}} & /^m \qw & \qw & \qw & \qw & \qw & \measure{\mbox{\text{update $y_{i+1}$}}} \cwx \\
& & & \dstick{\text{Repeat $\mathcal{O}(\sqrt{2^n})$ times}} & & & \text{}
\gategroup{1}{1}{2}{7}{1em}{--}
}
\end{equation*}
\caption{Quantum circuit of quantum minimization algorithm.}
\label{fig:circuitminimization}
\end{figure}
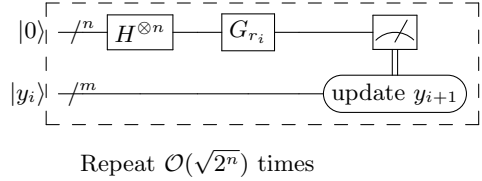

\noindent Now, we are going to present the generalized version of the algorithm. As an input the algorithm requires:
\begin{itemize}
\item Objective function $f(x): \{0,1\}^n \rightarrow R$,
\item Quantum oracle $O$ acting on $n$ input qubits in the first register.
\end{itemize}
Theoretically, a searched global optimum $x_{min}$ is measured with high probability at the end. The quantum circuit for that algorithm is presented in Fig.~(\ref{fig:circuitminimization}).
\\
\textbf{Set starting point} \\
Randomly select input $x_1$ and set $y_1=f(x_1)$ as a starting point of algorithm. \\\\
\textbf{Iteration (i): 1. Initialize} \\
Initialize the first register with $n$ input qubits and the second register with $m$ qubits to store threshold $y_i$. \\\\
\textbf{Iteration (i): 2. Apply Hadamard gates} \\
Apply Hadamard gates on the first register in order to put the first register into a superposition of all possible states representing $f(x)$ inputs. \\\\
\textbf{Iteration (i): 3. Apply Grover's algorithm} \\
Apply Grover's algorithm using quantum oracle $O$ and inversion around mean $r_i$ times. This is denoted in quantum circuit as $G_{r_i}$ gate. The measurement in Grover's algorithm is presented separately on quantum circuit. As mentioned earlier the value of $r_i$ is chosen randomly in order to avoid algorithm trapping. Denote output of Grover's algorithm by $\ket{x}$, convert that state to input $x$ and calculate value of function $y=f(x)$. \\\\
\textbf{Iteration (i): 4. Set new threshold} \\
If $y<y_i$ then set $x_{i+1}=x$ and $y_{i+1}=y$. In other case set value of $x$ and $y$ to previous values $x_{i+1}=x_i$ and $y_{i+1}=y_i$. This comparison operation can be executed on a classical computer. \\\\
\textbf{Convergence} \\
The convergence is achieved in order $\sqrt{2^n}$ iterations. The exact number of iterations is considered in Refs.~\cite{Durr,Ahuja}. Measurement of the first register gives a global minimum $x_{min}$ of the function with high probability. 

\subsubsection{Summary}
The quantum parallelism is achieved due to Grover's search, which simultaneously applies gates on all possible inputs in a superposition. The same fact allows this quantum algorithm to find global optimum as it is always searching through all possible inputs. The time complexity depends on a number of iterations in the main loop (around $\sqrt{2^n}$) and a number of inner loops $r_i$ in Grover's search algorithm. It turns out that for the purpose of minimization the numbers $r_i$ are substantially smaller than time complexity $\mathcal{O}(\sqrt{2^n})$ in standalone Grover's search algorithm \cite{Baritompa}. Thus, for very large $n$ only a number of iterations in the main loop is significant leading to overall $\mathcal{O}(\sqrt{2^n})$ time complexity of quantum minimization algorithm. This is a quadratic speed-up in comparison to classical optimization methods. The limitations of quantum minimization algorithm are the same as Grover's algorithm, the decoherence of quantum system can threaten the effectiveness of quantum algorithm destroying its quadratic speedup. As an alternative, one can consider a quantum approximate optimization algorithm (QAOA) \cite{Farhi} designed for noisy computers and used in training of quantum Boltzmann machine neural networks \cite{Verdon}. 

\subsection{Quantum Fourier transform algorithm}

\subsubsection{Preliminaries}
The quantum Fourier transform $QFT$ is a quantum analogue of the classical discrete Fourier transform $DFT$. The $DFT$ is effectively used in digital signal processing, image processing and data compression. The $DFT$ acts on a vector $x=(x_1,x_2,\dots,x_{N-1})^T$ and maps it to a new vector $y=(y_1,y_2,\dots,y_{N-1})^T$ in a following way:
\begin{align}
y_k=\frac{1}{\sqrt{N}}\sum_{j=0}^{N-1}x_je^{2\pi i \frac{kj}{N}}
\end{align}
In order to understand $DFT$ assume input signal consisting of true signal with frequency $f=10$ and amplitude $A=2$ described by equation:
\begin{align}
x_j^{(1)}=2\sin{(2\pi10j)}
\end{align} 
and a noise with high frequency $f=50$ and small amplitude $A=0.3$ described by equation
\begin{align}
x_j^{(2)}=0.3\sin{(2\pi50j)}
\end{align} 
The true signal and noise signal for $N=1000$ samples are presented in Fig.~(\ref{fig:inputsignal1}) and Fig.~(\ref{fig:inputsignal2}), respectively:
\begin{figure}[H]
\begin{minipage}{0.47\textwidth}	
\includegraphics[height=7.5cm,trim=.5cm .5cm .5cm .5cm,clip]{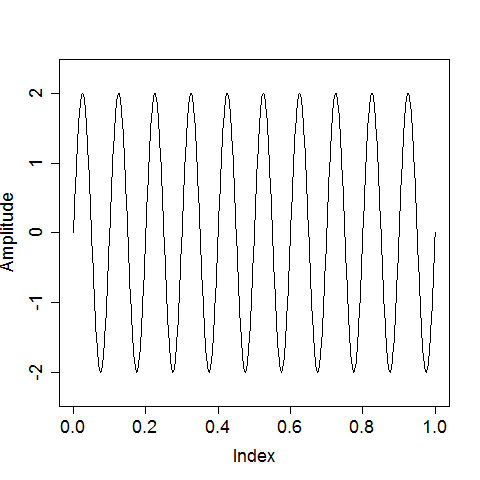}
\caption{Noise signal with $f=10$ and $A=2$}
\label{fig:inputsignal1}
\end{minipage}
\hfill
\begin{minipage}{0.47\textwidth}	
\includegraphics[height=7.5cm,trim=.5cm .5cm .5cm .5cm,clip]{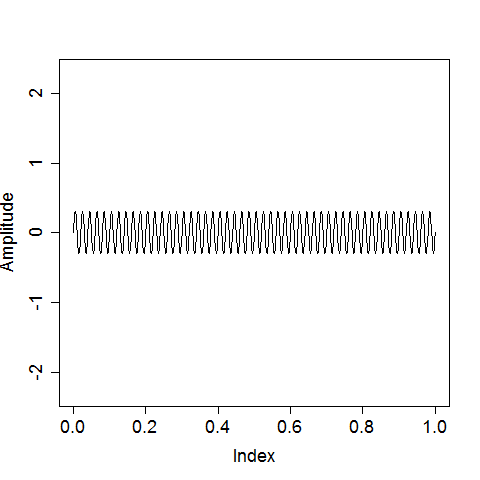}
\caption{Noise signal with $f=50$ and $A=0.3$}
\label{fig:inputsignal2}
\end{minipage}
\end{figure}
The overall input signal $x_j$ is illustrated in Fig.~(\ref{fig:inputsignal}) together with input signal in frequency domain (after application of $DFT$) in Fig.~(\ref{fig:dft}):
\begin{figure}[H]
\begin{minipage}{0.47\textwidth}
\includegraphics[height=7.5cm,trim=.5cm .5cm .5cm .5cm,clip]{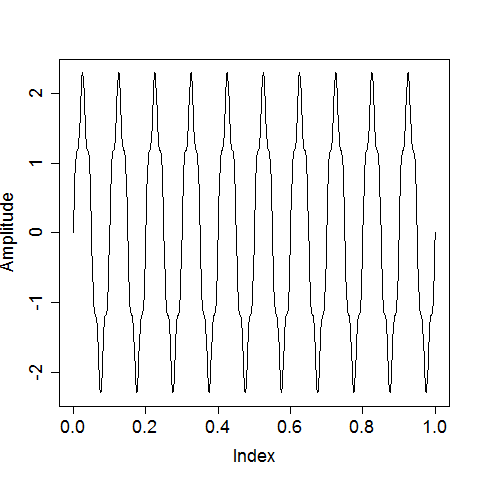}
\caption{Input signal $x_j$}
\label{fig:inputsignal}
\end{minipage}
\hfill
\begin{minipage}{0.47\textwidth}
\includegraphics[height=7.5cm,trim=.5cm .5cm .5cm .5cm,clip]{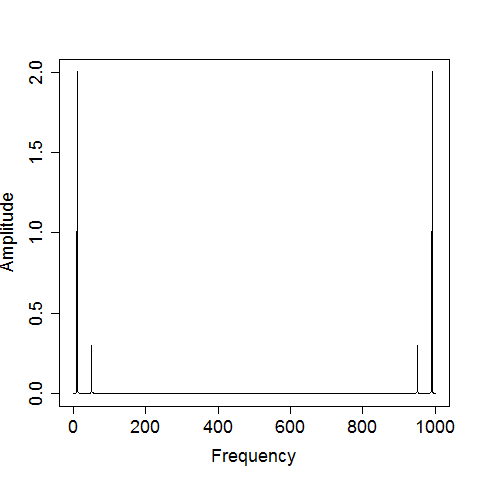}
\caption{Input signal in frequency domain $y_k$}
\label{fig:dft}
\end{minipage}
\end{figure}
We can see that exponentials $e^{2\pi i \frac{kj}{N}}$ work as frequency filters for different $k$ leaving peaks located in $k=10$ and $k=50$. The mirrored peaks at frequencies 950 and 990 are just the result of using complex numbers. Now, it is easier to reduce noise by discarding peak at frequency 50 (just set the coefficients $y_{50}$ and $y_{950}$ to zero) and apply inverse $DFT$. As a result we get true signal presented in Fig.~(\ref{fig:inputsignal1}). The quantum version of $DFT$ acts on amplitudes of a quantum state $\ket{\psi}=\sum_{j=0}^{N-1}a_j\ket{j}$ and maps them to new amplitudes of a quantum state $\ket{\phi}=\sum_{k=0}^{N-1}b_k\ket{k}$, where
\begin{align}
b_k=\frac{1}{\sqrt{N}}\sum_{j=0}^{N-1}a_je^{2\pi i \frac{kj}{N}}
\end{align}
or equivalently:
\begin{align}
\ket{j} \rightarrow \frac{1}{\sqrt{N}}\sum_{k=0}^{N-1}e^{2\pi i \frac{kj}{N}}\ket{k} \,.
\end{align}
The arising question is whether we can implement $QFT$ on a quantum computer as an effective quantum gate. In other words, we need to find an unitary operator that acts on a quantum state and do the $QFT$ operation. It turns out that such operator exists and is defined as:
\begin{align}
\mathcal{F}= \frac{1}{\sqrt{N}}\begin{bmatrix} 1 & 1 & 1 & \cdots & 1 \\ 1 & \omega & \omega^2 & \cdots & \omega ^{N-1} \\
1 & \omega^2 & \omega^4 & \cdots & \omega^{2(N-1)} \\ \vdots & \vdots & \vdots & & \vdots \\ 1 & \omega^{N-1} & \omega^{2(N-1)} & \cdots & \omega^{(N-1)(N-1)} 
\end{bmatrix}
\end{align}
where $\omega=e^{\frac{2\pi i}{N}}$.

\newpage
\subsubsection{Example}
\begin{figure}[h]
\begin{align*}
\Qcircuit @C=1em @R=1em {
\lstick{\ket{x_1}} & \qswap & \qw & \ctrl{1} & \gate{H} & \qw & \rstick{\ket{y_1}} \qw \\
\lstick{\ket{x_2}} & \qswap \qwx & \gate{H} & \gate{R_{\pi/2}} & \qw & \qw & \rstick{\ket{y_2}} \qw
}
\end{align*}
\caption{Quantum circuit of quantum Fourier transform algorithm presented in example.}
\label{fig:fourierexample}
\end{figure}
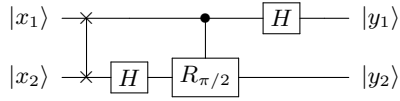

\noindent In that example, we are considering the 2-qubit state:
\begin{align}
\ket{\psi}=a_0\ket{0}+a_1\ket{1}+a_2\ket{2}+a_3\ket{3}=a_0\ket{00}+a_1\ket{01}+a_2\ket{10}+a_3\ket{11} \,.
\end{align}
The algorithm finds the Fourier transform: 
\begin{align} \label{eq:fourierPsi}
\mathcal{F}\ket{\psi}=\ket{\phi}=\sum_{k=0}^{3}b_k\ket{k}
\end{align}
with:
\begin{align} 
b_0&=\frac{1}{2}(a_0 + a_1 + a_2 + a_3) \nonumber \\
b_1&=\frac{1}{2}(a_0 + a_1e^{\frac{i\pi}{2}} + a_2e^{i\pi} + a_3e^{\frac{3i\pi}{2}} ) \nonumber \\
b_2&=\frac{1}{2}(a_0 + a_1e^{i\pi} + a_2e^{2i\pi} + a_3e^{3i\pi}) \nonumber \\
b_3&=\frac{1}{2}(a_0 + a_1e^{\frac{3i\pi}{2}} + a_2e^{3i\pi} + a_3e^{\frac{9i\pi}{2}} ) \,.
\end{align}
The detailed steps of the algorithm for that particular example are explained below. \\\\
\textbf{Apply Quantum Fourier Transform} \\
The 2-qubit $QFT$ gate can be represented by $N$-by-$N$ matrix where $N=2^2=4$:
\begin{align}
\mathcal{F}= \frac{1}{2}\begin{bmatrix} 1 & 1 & 1 & 1 \\ 1 & \omega & \omega^2 & \omega^{3} \\ 1 & \omega^2 & \omega^4 & \omega^{6} \\ 1 & \omega^{3} & \omega^{6} & \omega^{9} \end{bmatrix}
=\frac{1}{2}\begin{bmatrix} 1 & 1 & 1 & 1 \\ 1 & i & -1 & -i \\ 1 & -1 & 1 & -1 \\ 1 & -i & -1 & i \end{bmatrix}
\end{align}
as $\omega=e^{\frac{\pi i}{2}}=i$, $\omega^2=e^{\pi i}=-1$ and $\omega^4=e^{2\pi i}=1$ from Euler formula. By applying this gate on input state we get:
\begin{align}
\mathcal{F}\ket{\psi}= \frac{1}{2}\begin{bmatrix} 1 & 1 & 1 & 1 \\ 1 & i & -1 & -i \\ 1 & -1 & 1 & -1 \\ 1 & -i & -1 & i \end{bmatrix}
\begin{bmatrix} a_0 \\ a_1 \\ a_2 \\ a_3 \end{bmatrix} = 
\frac{1}{2}\begin{bmatrix} a_0 + a_1 + a_2 + a_3 \\ a_0 + ia_1 -a_2 -ia_3 \\ a_0 -a_1 + a_2 -a_3 \\ a_0 -ia_1 -a_2 + ia_3 \end{bmatrix} \,.
\end{align}
By careful comparison between this matrix and $\ket{\phi}$ state expressed by Eq.~(\ref{eq:fourierPsi}) we conclude that the desired result has been produced. \\\\
\textbf{Quantum circuit} \\
The $QFT$ gate can be represented by a set of simpler quantum gates. In this example, application of Hadamard gate $H$ on the first qubit, then controlled phase gate $R_{\frac{\pi}{2}}$ also on the first qubit (which is phase shift gate on the first qubit providing that the second qubit is in state $\ket{1}$) and lastly Hadamard gate on the second qubit reproduce the $QFT$ gate:
\begin{align}
\mathcal{F} &= (H\otimes I)(C\textnormal{-}R_{\frac{\pi}{2}})(I\otimes H)\textit{SWAP} \nonumber \\
&= \frac{1}{2} \begin{bmatrix} 1 & 0 & 1 & 0 \\ 0 & 1 & 0 & 1 \\ 1 & 0 & -1 & 0 \\ 0 & 1 & 0 & -1 \end{bmatrix} 
\begin{bmatrix} 1 & 0 & 0 & 0 \\ 0 & 1 & 0 & 0 \\ 0 & 0 & 1 & 0 \\ 0 & 0 & 0 & i \end{bmatrix} 
\begin{bmatrix} 1 & 1 & 0 & 0 \\ 1 & -1 & 0 & 0 \\ 0 & 0 & 1 & 1 \\ 0 & 0 & 1 & -1 \end{bmatrix}
\begin{bmatrix} 1 & 0 & 0 & 0 \\ 0 & 0 & 1 & 0 \\ 0 & 1 & 0 & 0 \\ 0 & 0 & 0 & 1 \end{bmatrix} \nonumber \\
&= \frac{1}{2}\begin{bmatrix} 1 & 1 & 1 & 1 \\ 1 & i & -1 & -i \\ 1 & -1 & 1 & -1 \\ 1 & -i & -1 & i \end{bmatrix} \,.
\end{align}
If we denote the first qubit by $x_1$ and the second qubit by $x_2$, then the quantum circuit representing matrix operations is presented in Fig.~(\ref{fig:fourierexample}).

\subsubsection{Algorithm}
\noindent An input to the algorithm is an arbitrary $n$-qubit state $\ket{\psi}=\sum_{j=0}^{N-1}a_j\ket{j}$ for which the Fourier transform is computed:
\begin{align} 
\mathcal{F}\ket{\psi}=\ket{\phi}=\sum_{k=0}^{N-1}b_k\ket{k}
\end{align}
with:
\begin{align}
b_k=\frac{1}{\sqrt{N}}\sum_{j=0}^{N-1}a_je^{2\pi i \frac{kj}{N}} \,.
\end{align}
\textbf{Apply Quantum Fourier Transform} \\
Apply $QFT$ gate:
\begin{align}
\mathcal{F}= \frac{1}{\sqrt{N}}\begin{bmatrix} 1 & 1 & 1 & \cdots & 1 \\ 1 & \omega & \omega^2 & \cdots & \omega ^{N-1} \\
1 & \omega^2 & \omega^4 & \cdots & \omega^{2(N-1)} \\ \vdots & \vdots & \vdots & & \vdots \\ 1 & \omega^{N-1} & \omega^{2(N-1)} & \cdots & \omega^{(N-1)(N-1)} 
\end{bmatrix}
\end{align}
where $\omega=e^{\frac{2\pi i}{N}}$. \\\\
\textbf{Quantum circuit} \\
The state $\ket{\psi}$ consisting of $n$ qubits can be represented by:
\begin{align}
\ket{\psi}=\ket{x_1,x_2,\dots,x_n}
\end{align}
with $x_i$ denoting state of $i$-th qubit. Similarly the $n$-qubit output state $\ket{\phi}$ can be represented by:
\begin{align}
\ket{\phi}=\ket{y_1,y_2,\dots,y_n} \,.
\end{align}
The input and output states $\ket{\psi}$ and $\ket{\phi}$ will be in a superposition, so that the qubits can be also entangled. The $QFT$ can be applied by a set of basic quantum gates: Hadamard gates and controlled phase shift gates. The reversed order of the input qubits represents multiple \textit{SWAP} gates applied at the beginning of the quantum circuit:
\begin{figure}[h]
\begin{equation*}
\Qcircuit @C=1em @R=.7em {
\lstick{\ket{x_n}} & \qw & \ctrl{6} & \ctrl{5} & \qw & \cdots & & \ctrl{3} & \qw & \cdots & & \ctrl{1} & \gate{H} & \rstick{\ket{y_1}} \qw \\
\lstick{\ket{x_{n-1}}} & \multigate{5}{\text{QFT}_{n-1}} & \qw & \qw & \qw & \cdots & & \qw & \qw & \cdots & & \gate{R_{\pi/2}} & \qw & \rstick{\ket{y_2}} \qw \\
\lstick{\vdots\ \ } & \pureghost{\text{QFT}_{n-1}} & & & & & & & & & & & & \rstick{\ \ \vdots} \\
\lstick{\ket{x_i}} & \ghost{\text{QFT}_{n-1}} & \qw & \qw & \qw & \cdots & & \gate{R_{\pi/2^{n-i}}} & \qw & \cdots & & \qw & \qw & \rstick{\ket{y_{n-i+1}}} \qw \\
\lstick{\vdots\ \ } & \pureghost{\text{QFT}_{n-1}} & & & & & & & & & & & & \rstick{\ \ \vdots} \\
\lstick{\ket{x_2}} & \ghost{\text{QFT}_{n-1}} & \qw & \gate{R_{\pi/2^{n-2}}} & \qw & \cdots & & \qw & \qw & \cdots & & \qw & \qw & \rstick{\ket{y_{n-1}}} \qw \\
\lstick{\ket{x_1}} & \ghost{\text{QFT}_{n-1}} & \gate{R_{\pi/2^{n-1}}} & \qw & \qw & \cdots & & \qw & \qw & \cdots & & \qw & \qw & \rstick{\ket{y_n}} \qw
}
\end{equation*}
\caption{Quantum circuit of quantum Fourier transform algorithm.}
\end{figure}
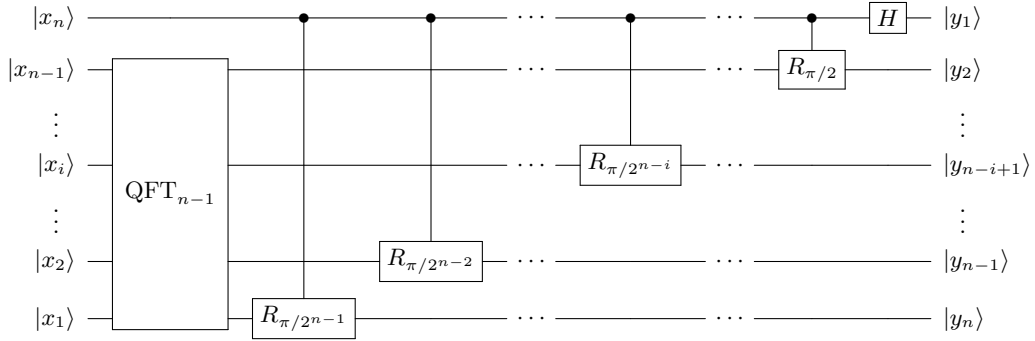

\subsubsection{Summary}
The quantum parallelism for that algorithm results from application of quantum gates acting on superposition (possibly entangled) of input quantum state. Moreover, the quantum inputs state can encode $2^n$ bits of classical data in just $n$ qubits. Analyzing the quantum circuit it can be concluded that the number of gates required for implementation of $QFT$ algorithm (omitting \textit{SWAP} gates at the beginning) is equal to $1+2+\dots+n=\frac{n(n-1)}{2}$, which corresponds to $\mathcal{O}(n^2)$ time complexity. The best algorithms improve that complexity to $\mathcal{O}(n\log{}n)$ \cite{Hales}. This is an exponential speed-up in comparison to $\mathcal{O}(n2^n)$ time complexity for classical discrete Fourier transform. 

\subsection{Quantum phase estimation algorithm}

\subsubsection{Preliminaries}
A common problem in applied mathematics is to find eigenvalue of a matrix given its eigenvector. In quantum computation this problem is rephrased into finding an eigenvalue $\lambda$ of eigenvector $\ket{\phi}$ satisfying $U\ket{\phi}=\lambda\ket{\phi}$ where $U$ is an unitary quantum gate. Due to the fact that $U$ is unitary, eigenvalue can be expressed as $\lambda=e^{2\pi i \theta}=\cos{(2\pi\theta)}+i\sin{(2\pi\theta)}$, with phase $0\leq \theta < 1$. The problem boils down to finding a phase $\theta$ with a particular level of precision. 

\newpage
\subsubsection{Example}
\noindent We have a unitary matrix:
\begin{align}
U= \begin{bmatrix} 1 & 0 \\ 0 & 1 \end{bmatrix} \,,
\end{align}
and the corresponding eigenvector:
\begin{align}
\ket{\phi}=\begin{bmatrix} 1 \\ 0 \end{bmatrix} \,.
\end{align}
We know that the eigenvalue for that problem is $\lambda=1$, which is equivalent to $\theta=0$. We apply quantum phase estimation algorithm to get the same result. The detailed steps of the algorithm for that particular example are explained below. \\\\
\textbf{Initialization} \\
We begin with initializing $n=1$ control qubit that will store a result:
\begin{align}
\ket{0}= \begin{bmatrix} 1 \\ 0 \end{bmatrix}
\end{align}
and $m=1$ qubit encoding eigenvector as:
\begin{align}
\ket{\phi}=\ket{0}=\begin{bmatrix} 1 \\ 0 \end{bmatrix} \,.
\end{align} \\\\
\textbf{Create superposition} \\
Then, we apply Hadamard gate on control qubit resulting in the state:
\begin{align}
(H\otimes I)\ket{0}\ket{0}&=\frac{1}{\sqrt{2}}
\begin{bmatrix} 1 & 0 & 1 & 0 \\ 0 & 1 & 0 & 1 \\ 1 & 0 & -1 & 0 \\ 0 & 1 & 0 & -1 \\ \end{bmatrix}
\begin{bmatrix} 1 \\ 0 \\ 0 \\ 0 \\ \end{bmatrix}
=\frac{1}{\sqrt{2}} \begin{bmatrix} 1 \\ 0 \\ 1 \\ 0 \\ \end{bmatrix} \nonumber \\
&= \frac{1}{\sqrt{2}}\left(\begin{bmatrix} 1 \\ 0 \end{bmatrix}+\begin{bmatrix} 0 \\ 1 \end{bmatrix}\right)\begin{bmatrix} 1 \\ 0 \end{bmatrix}
=\frac{1}{\sqrt{2}}(\ket{0}+\ket{1})\ket{0} \,.
\end{align} \\\\\\
\textbf{Apply controlled unitary gates} \\
We would like to apply controlled unitary gate \textit{C-U}, which applies gate $U$ on the eigenvector state only if control qubit is in state $\ket{1}$:
\begin{align}
\textit{C-U}\frac{1}{\sqrt{2}}(\ket{0}+\ket{1})\ket{0}=\frac{1}{\sqrt{2}}(\ket{0}\ket{0}+\ket{1}U\ket{0})=\frac{1}{\sqrt{2}}(\ket{0}+e^{2\pi i \theta}\ket{1})\ket{0} \,.
\end{align}
This can be rewritten as a following sum (the number 2 is kept in the denominator of exponential, because it would turn out to fit some pattern later):
\begin{align}
\frac{1}{\sqrt{2}}\sum_{k=0}^{1}e^{2\pi i \frac{k2\theta}{2}}\ket{k}\ket{0} \,.
\end{align}
Now, a result of the control qubit measurement does not depend on phase $\theta$ because the amplitude in exponential $e^{2\pi i k\frac{2\theta}{2}}$ is killed by its complex conjugate $e^{-2\pi i k\frac{2\theta}{2}}$ during calculation of probability. If only we can express the control qubit in a different basis which will be dependent on phase $\theta$. A closer look unveils that the control qubit state has a form of Fourier-transformed state:
\begin{align}
\ket{j}=\frac{1}{\sqrt{N}}\sum_{k=0}^{N-1}e^{2\pi i \frac{kj}{N}}\ket{k}
\end{align}
with $N=2$ and $j=2\theta$. The last step is to recover $\ket{j}=\ket{2\theta}$ by applying inverse operation to Fourier transform. \\\\
\textbf{Apply inverse quantum Fourier transform gate} \\
The inverse quantum Fourier transform creates a different basis:
\begin{align}
\mathcal{F}^{-1}\ket{k} = \frac{1}{\sqrt{N}}\sum_{j=0}^{N-1}e^{-2\pi i \frac{kj}{N}}\ket{j} \,.
\end{align}
For our example the inverse quantum Fourier transform gate for 2 qubits is expressed as:
\begin{align}
\mathcal{F}^{-1} =\frac{1}{\sqrt{2}} \begin{bmatrix} 1 & 1 \\ 1 & \omega^{-1} \end{bmatrix}
\end{align}
where $\omega=e^{\frac{2\pi i}{2}}=e^{\pi i}$. Thus:
\begin{align}
\mathcal{F}^{-1}\frac{1}{\sqrt{2}}\sum_{k=0}^{1}e^{2\pi i \frac{k2\theta}{2}}\ket{k}\ket{0}=
\frac{1}{2} \begin{bmatrix} 1 & 1 \\ 1 & e^{-\pi i} \end{bmatrix} \begin{bmatrix} 1 \\ e^{2\pi i \theta} \end{bmatrix} \begin{bmatrix} 1 \\ 0 \end{bmatrix}
=\frac{1}{2} \begin{bmatrix} 1 + e^{2\pi i \theta} \\ 1 + e^{\pi i} \end{bmatrix} \begin{bmatrix} 1 \\ 0 \end{bmatrix}
\end{align}
which is equivalent to:
\begin{align}
\frac{1}{2}\sum_{j=0}^{1}\sum_{k=0}^{1}e^{2\pi i \frac{k(2\theta-j)}{2}}\ket{j}\ket{0} \,.
\end{align} \\\\
\textbf{Measurement} \\
Measurement of the control qubit collapses the state of system to $\ket{2\theta}\ket{0}$. Note that the probability of measuring control qubit to be in the state $\ket{2\theta}$ is calculated as:
\begin{align}
P(\ket{2\theta}) &= \left|\frac{1}{2}\bra{2\theta}\sum_{j=0}^{1}\sum_{k=0}^{1}e^{2\pi i \frac{k(2\theta-j)}{2}}\ket{j}\right|^2 \nonumber \\
&=\left|\frac{1}{2}\sum_{k=0}^{1}e^{0}\braket{2\theta \vert 2\theta}\right|^2 \nonumber \\
&=\left|\frac{1+1}{2}\right|^2 \nonumber \\
&=1 \,.
\end{align} 
It means that measuring the control qubit always gives a correct $\theta$. In our example the qubit will be measured in $\ket{0}$ state as $\theta=0$. In cases where $\theta$ is not an integer or we operate on larger eigenvectors we need to increase a number of control qubits $n$ as well as a number of controlled quantum gates \textit{C-U}. General algorithm is presented in the next section.

\newpage
\subsubsection{Algorithm}
\begin{figure}[h]
\begin{equation*}
\Qcircuit @C=1em @R=1em @!R {
\lstick{\ket{0}} & \gate{H} & \qw & \qw & \qw & \qw & \cdots & & \ctrl{4} & \multigate{3}{\mathcal{F}^{-1}} & \qw & \meter & \\
\lstick{\ket{0}} & \gate{H} & \qw & \qw & \ctrl{3} & \qw & \cdots & & \qw & \ghost{\mathcal{F}^{-1}} & \qw & \meter & \\
\lstick{\vdots\ \ } & & & & & & & & & \pureghost{\mathcal{F}^{-1}} & & \vdots & \\
\lstick{\ket{0}} & \gate{H} & \qw & \ctrl{1} & \qw & \qw & \cdots & & \qw & \ghost{\mathcal{F}^{-1}} & \qw & \meter & \\
\lstick{\ket{\psi}} & /^m \qw & \qw & \gate{C\textnormal{-}U^{2^0}} & \gate{C\textnormal{-}U^{2^1}} & \qw & \cdots & & \gate{C\textnormal{-}U^{2^{n-1}}} & \qw & \qw
}
\end{equation*}
\caption{Quantum circuit of quantum phase estimation algorithm.}
\label{fig:circuitphase}
\end{figure}
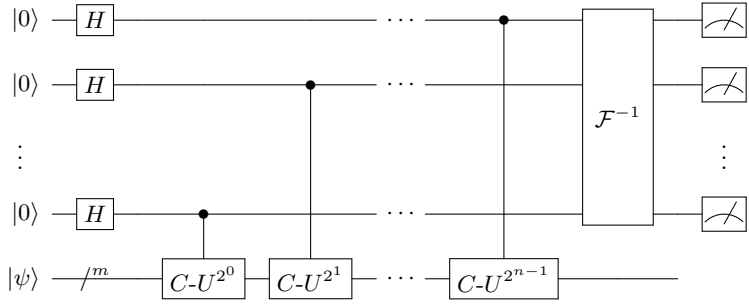

The general algorithm is represented by the quantum circuit in Fig.~(\ref{fig:circuitphase}). Similarly to the example presented in the previous subsection, we would require a quantum gate $U$ and its eigenvector $\ket{\phi}$ as an input. The algorithm reveals a phase $\theta$ allowing for calculation of an eigenvalue $\lambda=e^{2\pi i \theta}$. 
\\
\textbf{Initialization} \\
Initialize $n$ control qubits $\ket{0}$ and $m$ qubits encoding eigenvector $\ket{\phi}$. The initial state of the system is $\psi_0=\ket{0}^{\otimes n}\ket{\phi}$. \\\\
\textbf{Create superposition} \\
Apply $n$-bit Hadamard gate $H^{\otimes n}$ on $n$ control qubits resulting in the state:
\begin{align}
\ket{\psi_1}=\frac{1}{\sqrt{2^n}}(\ket{0}+\ket{1})^{\otimes n}\ket{\phi}
\end{align}
 \\
\textbf{Apply controlled unitary gates} \\
Controlled unitary gate \textit{C-U} applies gate $U$ on the state $\ket{\phi}$ only if the control qubit is in the state $\ket{1}$. Having $n$ control qubits we apply $n$ control unitary gates $\textit{C-U}^{2^j}$ with $0\leq j \leq n-1$ on the eigenstate $\ket{\phi}$. Noticing that:
\begin{align}
U^{2^j}\ket{\phi}=U^{2^j-1}U\ket{\phi}=U^{2^j-1}e^{2\pi i \theta}\ket{\phi}=e^{2\pi i 2^j \theta}\ket{\phi}
\end{align}
and moving the phases from eigenvector to control qubits we eventually end up with the state in a superposition:
\begin{align}
\ket{\psi_2}=\frac{1}{\sqrt{2^n}}\underbrace{\left(\ket{0}+e^{2\pi i 2^{n-1} \theta}\ket{1}\right)}_\text{qubit 1} \otimes \dots \otimes \underbrace{\left(\ket{0}+e^{2\pi i 2^{1} \theta}\ket{1}\right)}_\text{qubit n-1} \otimes \underbrace{\left(\ket{0}+e^{2\pi i 2^{0} \theta}\ket{1}\right)}_\text{qubit n}\otimes \ket{\phi} \,.
\label{eq:phaseestimation}
\end{align}
The Eq.~(\ref{eq:phaseestimation}) can be simplified by rewriting all $2^n$ $n$-qubits states $\{\ket{00\dots0},\ket{00\dots1},\dots,\ket{11\dots1}\}$ to states enumerated with an integer $\ket{k}$ made up from summing $2^j$'s in exponentials. As an example $e^{2\pi i 0 \theta}\ket{00\dots0}$ is rewritten as $e^{2\pi i 0 \theta}\ket{0}$, $e^{2\pi i 2^{0} \theta}\ket{0\dots01}$ is rewritten as $e^{2\pi i 1 \theta}\ket{1}$ and $e^{2\pi i (2^{1}+2^{0}) \theta}\ket{0\dots11}$ is rewritten as $e^{2\pi i 3 \theta}\ket{3}$. It can be generalized as:
\begin{align}
\ket{\psi_2}=\frac{1}{\sqrt{2^n}}\sum_{k=0}^{2^n-1}e^{2\pi i k\theta}\ket{k}\ket{\phi}=\frac{1}{\sqrt{2^n}}\sum_{k=0}^{2^n-1}e^{2\pi i \frac{k2^n\theta}{2^n}}\ket{k}\ket{\phi} \,.
\end{align} \\
\textbf{Apply inverse quantum Fourier transform gate} \\
Control qubits in the state $\ket{\psi_2}$ have form of a Fourier-transformed state. Applying inverse quantum Fourier transform gate $QFT^{-1}$ maps to a new basis where state containing phase gathers all amplitudes. The $QFT^{-1}$ gate applied on the previous state yields: 
\begin{align}
\ket{\psi_3} &= \frac{1}{2^n}\sum_{j=0}^{2^n-1}\sum_{k=0}^{2^n-1}e^{2\pi i k\theta}e^{-2\pi i \frac{k j}{2^n}}\ket{j}\ket{\phi} \nonumber \\
&= \frac{1}{2^n}\sum_{j=0}^{2^n-1}\sum_{k=0}^{2^n-1}e^{2\pi i \frac{k(2^n\theta-j)}{2^n}}\ket{j}\ket{\phi} \,.
\end{align} \\\\
\textbf{Phase approximation} \\
Due to the fact that states $\ket{j}$ can encode only discrete set of variables, whereas $\theta$ is a continuous variable in range $0\leq \theta < 1$ we approximate $2^n\theta=a + 2^n \delta$ where $a$ is the nearest integer of $2^n\theta$ and $\delta$ is rounding error with $0 \leq |2^n\delta| \leq \frac{1}{2}$. We can rewrite the previous state as:
\begin{align}
\ket{\psi_3}=\frac{1}{2^n}\sum_{j=0}^{2^n-1}\sum_{k=0}^{2^n-1}e^{2\pi i \frac{k(a-j)}{2^n}+2\pi i k \delta}\ket{j}\ket{\phi} \,.
\end{align} \\\\
\textbf{Measurement} \\
Now, measuring control qubits yields state $\ket{a}$ with probability (for simplicity we assume that $\delta=0$ i.e. there is no rounding error):
\begin{align}
P(\ket{a}) &=\left|\frac{1}{2^n}\bra{a}\sum_{j=0}^{2^n-1}\sum_{k=0}^{2^n-1}e^{2\pi i \frac{k(a-j)}{2^n}}\ket{j}\right|^2 \nonumber \\
&= \left|\frac{1}{2^n}\sum_{k=0}^{2^n-1}e^{2\pi i \frac{k(a-a)}{2^n}}\braket{a \vert a}\right|^2 \nonumber \\
&= \left|\frac{1}{2^n}\sum_{k=0}^{2^n-1}e^{0}\right|^2 \nonumber \\
&= 1 \,.
\end{align} 
It means that measuring control qubits always leaves them in the state $\ket{a}=\ket{2^n\theta}$ from which we can read off the phase and then calculate an eigenvalue. The state of the whole system after measurement is $\ket{2^n\theta}\ket{\phi}$. In case rounding error is not zero the state $\ket{a}$ will be yielded with probability:
\begin{align}
P(\ket{a}) &=\left|\frac{1}{2^n}\bra{a}\sum_{j=0}^{2^n-1}\sum_{k=0}^{2^n-1}e^{2\pi i \frac{k(a-j)}{2^n}+2\pi i k \delta}\ket{j}\right|^2 \nonumber \\
&= \left|\frac{1}{2^n}\sum_{k=0}^{2^n-1}e^{2\pi i k \delta}\right|^2 \nonumber \\
&=\frac{1}{2^{2n}}\left|\frac{1-e^{2\pi i 2^n \delta}}{1-e^{2\pi i \delta}}\right|^2 \,.
\end{align} 
It can be shown \cite{Clave} that $P(\ket{a}) \geq \frac{4}{\pi^2} \approx 0.405$ and converges to 1 as number of control qubits $n$ increases.

\subsubsection{Summary}
The quantum phase estimation algorithm allows us to calculate the eigenvalue for a given eigenvector of a quantum gate. It is used in quantum Principal Components Analysis algorithm described in section \ref{qPCA}. Assuming the \textit{C-U} gates are already prepared, the time complexity depends largely on the time complexity of quantum Fourier transform which is $\mathcal{O}(n\log{}n)$ for best-known algorithms \cite{Hales}. 

\section{Quantum machine learning algorithms}
This section provides an overview of selected unsupervised and supervised quantum machine learning algorithms, however there is also a method of scores extraction for quantum PCA algorithm proposed as well as a new cost function in feed-forward quantum neural networks is introduced. The unsupervised machine learning algorithms are quantum versions of $k$-means and $k$-median algorithms as well as quantum principal component analysis (qPCA). The supervised algorithms presented are quantum support vector machines (qSVM) and quantum neural networks (qNN). The quantum machine learning algorithms utilize the quantum algorithms described in the previous section as subroutines. The following names of subroutines have been assigned:
\begin{itemize}
\item \textit{GroverOptim}: Quantum minimization algorithm,
\item \textit{PhaseEstim}: Quantum phase estimation algorithm.
\end{itemize} 
Additional quantum subroutines have been introduced purely for the purpose of quantum machine learning algorithms:
\begin{itemize}
\item \textit{SwapTest},
\item \textit{DistCalc},
\item \textit{MedianCalc}.
\end{itemize} 
These are carefully explained in the following subsections. The mapping between quantum machine learning algorithms and quantum subroutines is presented (U - Used, O - Optional):
\renewcommand{\arraystretch}{1}
\begin{table}[H]
\caption{Quantum machine learning algorithms and quantum subroutines mapping}
\vspace*{1em}
\centering
\begin{tabular}{ r | c | c | c | c | c | }
\multicolumn{1}{r}{}
& \multicolumn{1}{c}{\textit{GroverOptim}}
& \multicolumn{1}{c}{\textit{PhaseEstim}}
& \multicolumn{1}{c}{\textit{SwapTest}}
& \multicolumn{1}{c}{\textit{DistCalc}}
& \multicolumn{1}{c}{\textit{MedianCalc}} \\
\cline{2-6}
q $k$-means & U & & U & U & \\
\cline{2-6}
q $k$-medians & U & & U & U & U \\
\cline{2-6}
qSVM & U & & & & \\
\cline{2-6}
qPCA & & U & O & & \\
\cline{2-6}
qNN & & & & & \\
\cline{2-6}
\end{tabular}
\end{table}

\subsection{Quantum $k$-means algorithm}
\subsubsection{Preliminaries}
Classical $k$-means algorithm belongs to the family of unsupervised machine learning algorithms. It aims to classify data to $k$ clusters based on an unlabeled set of training vectors. The training vectors are reassigned to the nearest centroid in each iteration and then a new centroid is calculated averaging the vectors belonging to the current cluster. The time complexity $\mathcal{O}(NM)$ of the classical algorithm using Lloyd’s version of the algorithm is linearly dependent on a number of features $N$ and a number of training examples $M$ \cite{Manning}. The most resource consuming operation for $k$-means algorithm is calculation of a distance between vectors and we are expecting to gain a speed-up by retrieving the distance using a quantum computer. The quantum $k$-means algorithm turns out to provide an exponential speed-up for very large dimension of a training vector. The algorithm is based on two new quantum subroutines calculating a distance between vectors: \textit{SwapTest} and \textit{DistCalc} as well as the quantum subroutine \textit{GroverOptim} already described in section \ref{qGroverOptim} and assigning vector to the closest centroid of cluster. 

\subsubsection{Quantum subroutine: \textit{SwapTest}} \label{QSwapTest}
\begin{figure}[h]
\begin{equation*}
\Qcircuit @C=1em @R=1em @!R {
\lstick{\ket{0}} & \gate{H} & \ctrl{2} & \gate{H} & \qw & \meter \\
\lstick{\ket{a}} & \qw & \qswap & \qw & \qw \\
\lstick{\ket{b}} & \qw & \qswap & \qw & \qw \\
}
\end{equation*}
\caption{Quantum circuit of \textit{SwapTest} quantum subroutine.}
\label{fig:circuitswap}
\end{figure}
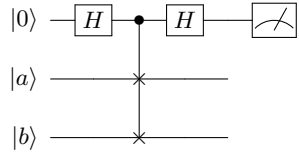

The subroutine \textit{SwapTest} is a simple quantum routine expressing overlap of two states $\braket{a \vert b}$ in terms of measurement probability of control qubit being in state $\ket{0}$. The overlap is a measure of similarity between quantum states and it will be used in calculation of a distance between classical vectors in \textit{DistCalc} subroutine. The algorithm has been firstly used in Ref.~\cite{aimur}. The quantum circuit of the algorithm is presented in Fig.~(\ref{fig:circuitswap}). 
\newpage
\noindent The steps of the subroutine are presented below. \\\\
\textbf{Initialize} \\
Initialize two states $\ket{a}$ and $\ket{b}$ as well as a control qubit $\vert 0 \rangle$ resulting in the state:
\begin{align}
\ket{\psi_0}=\ket{0,a,b} \,.
\end{align}
The states $\ket{a}$ and $\ket{b}$ consist of $n$ qubits each. 
\\\\
\textbf{Apply Hadamard gate} \\
Apply Hadamard gate on the control qubit resulting in a superposition:
\begin{align}
\ket{\psi_1}=(H\otimes I^{\otimes n}\otimes I^{\otimes n})\ket{\psi_0}=\frac{1}{\sqrt{2}}(\ket{0,a,b}+\ket{1,a,b}) \,.
\end{align} \\
\textbf{Apply \textit{SWAP} gate} \\
Apply controlled \textit{SWAP} gate on $\ket{a}$ and $\ket{b}$ states which swaps $a$ and $b$ providing that the control qubit is in state $\ket{1}$. As a result:
\begin{align}
\ket{\psi_2}=\frac{1}{\sqrt{2}}(\ket{0,a,b}+\ket{1,b,a}) \,.
\end{align} \\
\textbf{Apply Hadamard gate} \\
Apply second Hadamard gate on the control qubit resulting in the state:
\begin{align}
\ket{\psi_3}=\frac{1}{2}\ket{0}(\ket{a,b}+\ket{b,a})+\frac{1}{2}\ket{1}(\ket{a,b}-\ket{b,a}) \,.
\end{align} \\
\textbf{Measurement} \\
Measure the control qubit. The probability of measuring control qubit being in state $\ket{0}$ is given by:
\begin{align}
P(\ket{0}) &=|\frac{1}{2}\braket{0 \vert 0}(\ket{a,b}+\ket{b,a})+\frac{1}{2}\braket{0 \vert 1}(\ket{a,b}-\ket{b,a})|^2 \nonumber \\
&=\frac{1}{4}|(\ket{a,b}+\ket{b,a})|^2 \nonumber \\
&=\frac{1}{4}(\braket{b \vert b} \braket{a \vert a}+\braket{b \vert a} \braket{a \vert b}+\braket{a \vert b} \braket{b \vert a}+\braket{a \vert a} \braket{b \vert b}) \nonumber \\
&=\frac{1}{2}+\frac{1}{2}|\langle a \vert b \rangle|^2
\end{align}
Thus, we successfully linked an overlap $\braket{a \vert b}$ with measurement probability of the control qubit in the final quantum state. The probability $P(\ket{0})=0.5$ means that the states $\ket{a}$ and $\ket{b}$ are orthogonal, whereas the probability $P(\ket{0})=1$ indicates that the states are identical. The subroutine should be repeated several times to obtain a good estimator of probability. The advantage of using swap test is that the states $\ket{a}$ and $\ket{b}$ can be unknown before procedure and simple measurement is performed on the control qubit which has two eigenstates. The time complexity is negligible as the procedure does not depend on a number of qubits representing the input states, however we may pay attention to the preparation time of identical copies of $\ket{a}$ and $\ket{b}$. 

\subsubsection{Quantum subroutine: \textit{DistCalc}} \label{QDistCalc}
With subroutine \textit{SwapTest} presented we can move on to the algorithm retrieving the Euclidean distance $|a-b|^2$ between two real valued vectors $a$ and $b$. The algorithm was described by Lloyd, Mohseni and Rebentrost \cite{Lloyd}. \\\\
\textbf{Representation of classical data as quantum states} \\
The classical information in vector $a$ is encoded as \cite{Niemann,Vartiainen}:
\begin{align}
|a|^{-1}a \rightarrow \ket{a} =\sum_{i=1}^N |a|^{-1}a_i\ket{i} \,. \label{eq:information}
\end{align}
Norm of quantum state is normalized with this definition $\braket{a\vert a}=|a|^{-2}a^2=1$, which leads to the correct definition of a quantum state. The $N$ dimensional training vector can be translated into $n=\log_2{N}$ qubits. As an example the vector with eight features can be stored in three qubits containing $2*2*2=8$ entries to express vector components $a_i$ as probability amplitudes. \\\\
\textbf{Initialize} \\
Initialize two quantum states:
\begin{align}
\ket{\psi}&=\frac{1}{\sqrt{2}}(\ket{0,a}+\ket{1,b}) \,, \\
\ket{\phi}&=\frac{1}{\sqrt{Z}}(|a|\ket{ 0}+|b|\ket{1})
\end{align}
with $Z=|a|^2+|b|^2$ \,. \\\\
\textbf{Use quantum subroutine \textit{SwapTest}} \\
Evaluate an overlap $\braket{\phi \vert \psi}$ using subroutine \textit{SwapTest}. \\\\
\textbf{Calculate distance} \\
Calculate Euclidean distance by noting that:
\begin{align}
|a-b|^2=2Z|\braket{\phi \vert \psi}|^2 \,.
\end{align}
This holds true, because:
\begin{align}
\braket{\phi \vert \psi}=\frac{1}{\sqrt{2Z}}(|a|\ket{a}-|b|\ket{b})
\end{align} 
and using the fact how the classical information has been prepared in Eq.~(\ref{eq:information}) the inner product could be expressed as:
\begin{align}
\braket{\phi \vert \psi}=\frac{1}{\sqrt{2Z}}(a-b)
\end{align}
so that $|\braket{\phi \vert \psi}|^2=\frac{1}{2Z}|a-b|^2$. Moreover, using this algorithm it is also easy to calculate the inner product between two vectors noticing that:
\begin{align}
a^Tb=\frac{1}{2}(|a|^2+|b|^2-|a-b|^2) \,.
\end{align}
To summarize, in \textit{DistCalc} subroutine we prepare two states, apply subroutine \textit{SwapTest} and repeat that procedure to obtain an acceptable estimate of probability. Providing that the states are already prepared the subroutine \textit{SwapTest} does not depend on the size of a feature vector. The preparation of states in subroutine \textit{DistCalc} is proved to has $\mathcal{O}(\log{N})$ time complexity \cite{Lloyd}, which makes sense as the classical information is encoded in $n=log_2{N}$ qubits and we expect the time complexity to be proportional. The classical algorithms require $\mathcal{O}(N)$ to calculate Eucidean distance between two vectors, thus there is an exponential speed-up. The measurement during swap test is causing the deconherence of input states, thus the quantum memory should contain multiple copies of input states, however it does not change time complexity for large $N$, as the number of states preparation will be always much smaller than $N$. 

\subsubsection{Algorithm}
Now, having two quantum subroutines \textit{SwapTest} and \textit{DistCalc} required to calculate distance and quantum optimization subroutine \textit{GroverOptim} to choose the closest centroid of cluster we are ready to put down the steps of quantum $k$-means algorithm. \\\\
\textbf{Initialize} \\
Choose number of $k$ clusters and randomly initialize $k$ cluster centroids $\mu_1,\mu_2,\dots,\mu_k \in R^N$. These initializations can by done by any methods used in classical $k$-means algorithm. \\\\
\textbf{Main Loop (s): Do until convergence is reached} 
\begin{quote}
\textbf{Inner Loop (i): Choose closest cluster centroid} \\
Loop over training examples $i=\{1,\dots,M\}$ and for each training example $x^{(i)}$ use quantum subroutine \textit{DistCalc} to calculate $k$ distances $\|x^{(i)}-\mu_k \|$ to each cluster centroids and then use quantum subroutine \textit{GroverOptim} to choose index $c^{(i)}=\{1,\dots,k\}$ of cluster centroid that minimize the distance between training example and cluster centroid:
\begin{align}
c^{(i)}:=arg\left\{ \min_{k} \|x^{(i)}-\mu_k \|^2 \right\} \,.
\end{align}
\textbf{Inner Loop (j): Calculate new cluster centroids } \\
Loop over clusters $j=\{1,\dots,k\}$ and calculate mean $\mu_j$ of points assigned to cluster $j$. The calculated $\mu_j$ becomes new cluster centroid. Specifically, for each centroid $j$ we set:
\begin{align}
\mu_j=\frac{1}{|C_j|}\sum_{i\in C_j}x^{(i)}
\end{align}
where $|C_j|$ is the number of training vectors belonging to cluster $j$ and $C_j$ is the set of these vectors. For example, imagine that there are four training examples belonging to cluster $j=2$: $C_2=\{x^{(2)},x^{(6)},x^{(10)},x^{(11)}\}$. The new cluster is calculated as $\mu_2=\frac{1}{4}(x^{(2)}+x^{(6)}+x^{(10)}+x^{(11)})$.
This operation is executed on a classical computer. 
\end{quote}
\textbf{Convergence}: \\
Convergence is reached when subsequent iterations of algorithm does not change the location of cluster centroids. Mathematically speaking, the algorithm converges when for each cluster $j=\{1,\dots,k\}$ the distance between cluster centroids from iteration $s+1$ and the previous iteration $s$ is lower than some arbitrary threshold $\eta$:
\begin{align}
\|\mu_j^{s+1}-\mu_j^{s}\| < \eta
\end{align}

\subsubsection{Summary}
The quantum $k$-means algorithm provides an exponential speed-up in comparison to classical $k$-means algorithm. The Lloyd's version of $k$-means algorithm executed on a classical computer has time complexity $\mathcal{O}(MNk)$. This just corresponds to the fact that for each training example we need to calculate distances between $N$ dimensional feature vectors and $k$ clusters. The number of algorithm iterations is assumed to be negligible in case we deal with large $M$, $N$ and $k$ (of course the algorithm can do many iterations before it converges, but we assume that it is not the main source of time complexity). The quantum parallelism is achieved thanks to the clever calculation of distances between large dimensional vectors. Thus, the time complexity of the quantum algorithm is $\mathcal{O}(Mk\log(N))$, which for very large $N$ is an exponential speed-up. Note that in the following time complexities the minimization procedure is not taken into account, so for a large number of clusters quantum minimization subroutine provides even more speed-up providing that the quantum oracles for Grover's algorithm are prepared efficiently. However, for a small number of clusters it could be advisable to use classical optimization algorithms, as D{\"u}rr and H{\o}yer minimization algorithm does it best for a large number of objective function inputs. We should also remember about limitations of quantum minimization algorithm in the presence of quantum noise. Further improvements of the algorithm exist, which uses a quantum adiabatic algorithm and quantum output \cite{Lloyd}. 

\subsection{Quantum $k$-medians algorithm}
\subsubsection{Preliminaries}
The $k$-medians algorithm is similar to $k$-means algorithm, however instead of setting new cluster centroids by calculating mean, a median is evaluated. The $k$-means algorithm has a disadvantage that in some cases, the calculated mean may lie outside the desired region, whereas cluster centroid calculated as median always belongs to a training vectors set. It has even further implications for quantum algorithms, as the calculation of mean is executed on a classical computer, whereas the calculation of median can be implemented using \textit{MedianCalc} subroutine \cite{aimur2}. The quantum version of $k$-medians algorithm requires the same quantum subroutines as quantum $k$-means algorithm: \textit{SwapTest}, \textit{DistCalc} and \textit{GroverOptim} plus an additional \textit{MedianCalc}, which also uses \textit{GroverOptim}. The quantum minimization algorithm \textit{GroverOptim} is optional to use in case of $k$-means algorithm, whereas for $k$-medians it may be used to choose closest centroids and must be used to calculate median. 

\subsubsection{Quantum subroutine: \textit{MedianCalc}} \label{QMedianCalc}
The quantum subroutine to calculate median $Q$ from set of $N$ dimensional points $\{a_1,a_2,\dots,a_m\}$ have been introduced in \cite{aimur2}. By definition, the median is a point within the set $\{a_1,a_2,\dots,a_m\}$ whose distance to all other points is minimum. As a consequence, in order to calculate median we must evaluate the following sum for each $a_i$:
\begin{align}
S_i = \sum_{j=1}^m\|a_i-a_j\| 
\end{align}
and take a minimum. The index $i_{min}$ for the smallest $S_i$ denotes a median $Q=a_{i_{min}}$. The inputs to the algorithm are listed below: 
\begin{itemize}
\item Set of $N$ dimensional points $\{a_1,a_2,\dots,a_m\}$,
\item Quantum oracle $O$ used in \textit{GroverOptim} subroutine.
\end{itemize}
Below, we are giving the description of the subroutine. \\\\
\textbf{Calculation of distances} \\
Use \textit{DistCalc} subroutine to calculate all required distances for sums $\{S_1,\dots,S_m\}$. \\\\
\textbf{Use quantum subroutine \textit{GroverOptim}} \\
Use quantum minimization subroutine to choose smallest $S_i$. \\\\
\textbf{Median} \\
The index $i_{min}$ for the smallest $S_i$ will define median $Q=a_{i_{min}}$. Generally, the operation of choosing minimal $S_i$ executed on classical computer has time complexity of $\mathcal{O}(m)$. Assuming that a black-box quantum oracle $O$ has been already provided, the \textit{GroverOptim} subroutine can be used as it is a minimization problem. The optimization operation on quantum computer has time complexity $\mathcal{O}(\sqrt{m})$. There is also a gain in calculation of Euclidean distance between vectors using quantum subroutine \textit{DistCalc}. Assuming that dimension of feature vector $N$ is much larger than a number of vectors $m$ we gain an additional exponential speed-up as presented in section \ref{QDistCalc}.

\subsubsection{Algorithm}
\noindent The algorithm is similar to quantum $k$-means algorithm.  \\\\
\textbf{Initialize} \\
Choose number of $k$ clusters and randomly generate $k$ cluster centroids $\mu_1,\mu_2,\dots,\mu_k \in R^N$ from training examples set $x^{(1)},x^{(2)},\dots,x^{(M)}$ \,. \\\\
\newpage
\textbf{Main Loop (s): Do until convergence is reached} 
\begin{quote}
\textbf{Inner Loop (i): Choose closest cluster centroid} \\
Loop over training examples $i=\{1,\dots,M\}$ and for each training example $x^{(i)}$ use quantum subroutine \textit{DistCalc} to calculate $k$ distances $\|x^{(i)}-\mu_k \|$ to each cluster centroids and then use quantum subroutine \textit{GroverOptim} to choose index $c^{(i)}=\{1,\dots,k\}$ of cluster centroid that minimize the distance between training example and cluster centroid:
\begin{align}
c^{(i)}:=arg\left\{ \min_{k} \|x^{(i)}-\mu_k \|^2 \right\} \,.
\end{align}
\textbf{Inner Loop (j): Calculate new cluster centroids } \\
Loop over clusters $j=\{1,\dots,k\}$ and calculate median $\mu_j$ of points assigned to cluster $j$ using quantum subroutine \textit{MedianCalc}.
\end{quote}
\textbf{Convergence}: \\
Convergence is reached when subsequent iterations of algorithm does not change a location of cluster centroids. For $k$-medians algorithm it just means that each cluster $j=\{1,\dots,k\}$ in iteration $s+1$ is the same as in the previous iteration $s$:
\begin{align}
\mu_j^{s+1}=\mu_j^{s} \,.
\end{align}

\subsubsection{Summary}
The time complexity of quantum $k$-median algorithm is similar to quantum $k$-mean algorithm except for the time complexity difference between calculation of medians and means. In comparison to its classical counterpart, there is an exponential speed up coming from the same source as in quantum $k$-means algorithm and additional quadratic gain coming from the calculation of median on a quantum computer. The quantum parallelism of \textit{MedianCalc} is achieved by application of Grover's search in quantum minimization subroutine. Although $k$-median algorithm has the advantage of having clusters always in the desirable region, the calculation of median on a quantum computer can be longer than the simple calculation of mean on the classical computer due to the exhaustive evaluation of distances. This doubt can be disregarded if we assume that the distances are already given to the quantum $k$-median algorithm. This is a very interesting assumption which is based on the fact that instead of giving a set of training examples to $k$-median procedure, we can just calculate distances between all training examples and take it as the only input to the algorithm. Thus, the $k$-median algorithm, in this case, is more attractive than $k$-means algorithm if we can only obtain distances instead of vectors as a training data. 

\subsection{Quantum support vector machines} 
\subsubsection{Preliminaries}
The support vector machines algorithm (SVM) is one of supervised machine learning algorithms used for linear discrimination problems. The algorithm is based on finding a hyperplane that discriminates between two classes of feature vectors and is used as a decision boundary for future data classification. The SVM is formulated as maximizing the distance between the hyperplane and closest data points called support vectors. The objective function could be convex or non-convex depending on the kernel used in SVM algorithm. The non-convex functions tend to converge to a local optimum, thus the classical SVM balances between optimization effectiveness and the prediction accuracy. Quantum version of support vector machines using minimization quantum subroutine \textit{GroverOptim} and provides convergence to a global optimum for non-convex cost function \cite{anguita}. 

\subsubsection{Algorithm}
For simplicity, assume linear support vector machines with hyperplanes defined by:
\begin{align}
\theta^Tx-b=\pm1
\end{align}
and a distance between two hyperplanes expressed by $\|\frac{2}{\theta}\|$. Note that $x$ and $\theta$ are vectors, whereas $b$ is a numeric constant. Minimizing $\theta$ leads to a maximum margin and adding constraints would ensure that the margin correctly classifies the data. This can be represented by the following optimization problem:
\begin{align}
\min_{\theta,b}\frac{1}{2}\|\theta\|^2
\end{align}
subject to a constraint:
\begin{align}
y^{(i)}(\theta^Tx^{(i)}-b) \geq 1
\end{align}
for all training examples $i=\{1,\dots,M\}$ and $y^{(i)}=\{-1,1\}$. The constraint can be incorporated into an objective function using Lagrange multipliers $\alpha^{(i)}$, which results in following formulation of the problem:
\begin{align}
\min_{\theta,b}\max_{\alpha^{(i)} \geq 0}\left(\frac{1}{2}\|\theta\|^2-\sum_{i=1}^M\left[\alpha^{(i)}(\theta^Tx^{(i)}-b)-1\right]\right) \,.
\end{align}
Note, that the only non-zero $\alpha^{(i)}$ will corresponds to $x^{(i)}$ being support vectors. In order to solve the maximization of objective function $F$ with respect to $\alpha^{(i)}$, we set the following derivatives to zero:
\begin{align}
\frac{\partial{F}}{\partial{\theta^{(i)}}}&=\theta^{(i)}-\alpha^{(i)}y^{(i)}x^{(i)}=0 \nonumber \\
\frac{\partial{F}}{\partial{b}}&=\sum_{i=1}^M\alpha^{(i)}y^{(i)}=0 \,.
\end{align}
As a consequence, we can express weights as:
\begin{align}
\theta=\sum_{i=1}^M\alpha^{(i)}y^{(i)}x^{(i)}
\end{align}
and the dual problem as:
\begin{align}
\min_{\alpha^{(i)}}\left\{\frac{1}{2}\sum_{i,j}\alpha^{(i)}\alpha^{(j)}y^{(i)}y^{(j)}(x^{(i)})^Tx^{(j)}-\sum_{i=1}^M\alpha^{(i)}\right\}
\end{align}
providing that:
\begin{align}
\alpha^{(i)} \geq 0 
\end{align}
for each training example $i=1,\dots,M$ and:
\begin{align}
\sum_{i=1}^M\alpha^{(i)}y^{(i)}=0 \,.
\end{align}
The optimization problem can be generalized to an arbitrary kernel function $K(x^{(i)},x^{(j)})$ introducing the non-linearity to the problem. The dot product in the previous dual problem is replaced with a kernel function:
\begin{align}
\min_{\alpha^{(i)}}\left\{\frac{1}{2}\sum_{i,j}\alpha^{(i)}\alpha^{(j)}y^{(i)}y^{(j)}K(x^{(i)},x^{(j)})-\sum_{i=1}^M\alpha^{(i)}\right\} \,.
\end{align}
As an example, we can take Gaussian kernel function expressed as:
\begin{align}
K(x^{(i)},x^{(j)})=\exp{\left(-\gamma\|x^{(i)}-x^{(j)}\|^2\right)}
\end{align}
so that additional calculation of Euclidean distances is required. Now, we are going to give the detailed description of each algorithm's step. \\\\
\textbf{Initialize the free parameters and kernel function} \\
Initialize the value of any free parameters that are used in a kernel function. Choose an appropriate kernel function for the problem and calculate a kernel matrix. \\\\
\textbf{Representation of classical data and parameters as quantum states} \\
This step implies discretization of objective function and writing its components to qubits. The classical information can be represented as binary string:
\begin{align}
x \rightarrow a=(a_1,a_2,\dots,a_k)^T
\end{align}
with $a_i=\{0,1\}$ for $i=1\dots,k$. Then the binary strings can be directly translated into $k$ qubit quantum state:
\begin{align}
\ket{a_1 a_2 \dots a_k}
\end{align}
from a $2^k$ dimensional Hilbert space spanned by basis $\{\ket{00\dots0},\ket{10\dots0},\dots,\ket{11\dots1}\}$. \\\\
\textbf{Scanning objective function space} \\
The quantum minimization subroutine is searching through the objective function space to find an optimal set of $\alpha^{(i)}$, which solves for paramters $\theta$ and $b$. The subroutine \textit{GroverOptim} creates the superposition of all possible inputs to a quantum oracle $O$ representing objective function and finds a global minimum for the SVM optimization problem. The measurement in subroutine \textit{GroverOptim} reveals the solution with high probability.

\subsubsection{Summary}
The quantum subroutine \textit{GroverOptim} decreases the time complexity quadratically from $\mathcal{O}(N)$ to $\mathcal{O}(\sqrt{N})$ and provides a global minimum of the formulated SVM optimization problem. Still, the most time-consuming part of SVM algorithm is the calculation of a kernel matrix, which has the time complexity of $\mathcal{O}(M^2N)$ on a classical computer. The limitations of the algorithm are the same as for quantum subroutine \textit{GroverOptim}, i.e. the algorithm would not provide satisfying results in the presence of quantum noise. Although, we assume that a quantum oracle representing objective function of SVM is an input to the algorithm, we should not ignore that building such an oracle could be complex or impossible. An unconstrained exponential speed-up is presented in another quantum algorithm for SVM presented in Ref.~\cite{rebentrost}. 

\subsection{Quantum principal component analysis} \label{qPCA}
\subsubsection{Preliminaries}
Principal component analysis (PCA) is widely used dimensionality reduction procedure. It takes $N$ dimensional feature vectors (possibly correlated) from a training set, applies an orthonormal transformation and outputs $R$ dimensional data. The compressed data can be further used in other machine learning algorithms allowing to draw the same conclusions as on the full data, however executing algorithms much faster as in some cases $R<<N$. The PCA relies on the eigendecomposition of the correlation (or covariance) matrix, thus a quantum speed-up should be obtained in that area. The quantum phase estimation subroutine \textit{PhaseEstim} deals with eigenvectors and eigenvalues and indeed is used by quantum principal component analysis (qPCA). The qPCA algorithm is exponentially faster than any known classical algorithm and has been presented by Lloyd, Mohseni and Rebentrost \cite{LloydPCA}. The algorithm uses the quantum subroutine \textit{PhaseEstim} on an exponent of density matrix $\rho$ and produces the quantum states containing all eigenvectors and eigenvalues of the density matrix $\rho$. We will see that a density matrix $\rho$ is equivalent to covariance/correlation matrix. 

\subsubsection{Algorithm}
\noindent \textbf{Demean and normalization of classical data} \\
Firstly, a set of $N$ dimensional training vectors should be demeaned in order to properly use it in qPCA. For each training example $x^{(i)}$ with $i=\{1,\dots,M\}$ we subtract the $N$ dimensional vector with means $\bar{x}$:
\begin{align}
x^{(i)} &\rightarrow x^{(i)}-\bar{x} \nonumber \\
\bar{x} &= \frac{1}{M}\sum_{i=1}^Mx^{(i)} \,.
\end{align}
Secondly, the data should be normalized to build proper quantum states. We divide each training example $x^{(i)}$ by a norm of vector:
\begin{align}
x^{(i)} \rightarrow \left|x^{(i)}\right|^{-1}x^{(i)}
\end{align}
with the norm defined as:
\begin{align}
|x|=\sqrt{\sum_{k=1}^N x_k^2} \,.
\end{align}
We can also standardize the data in order to get correlation matrix, but this is optional. \\\\
\textbf{Representation of classical data as quantum states} \\
The components $x_k$ with $k=\{1,\dots,N\}$ of classical training vector $x$ are encoded as amplitudes of a quantum state \cite{Niemann,Vartiainen}:
\begin{align}
x \rightarrow \ket{x} =\sum_{k=1}^N x_k\ket{k} \,.
\end{align}
Due to the previous normalization of classical data, the quantum state is correctly defined with probabilities summing up to 1: $\braket{x\vert x}=x^2=1$. The $N$ dimensional training vector can be translated into $n=\log_2{N}$ qubits. \\\\
\textbf{Representation of covariance/correlation matrix as density matrix} \\
Having all training examples encoded as states, we are building a mixed state defined by the following density matrix:
\begin{align}
\rho = \frac{1}{M} \sum_{i=1}^M \ket{x^{(i)}} \bra{x^{(i)}} \,.
\end{align}
The tensor product $\ket{x^{(i)}} \bra{x^{(i)}}$ is written as:
\begin{align}
\ket{x^{(i)}} \bra{x^{(i)}} = \sum_{k=1}^N \sum_{m=1}^N x_k^{(i)} x_m^{(i)}\ket{k}\bra{m}
\end{align}
which in a matrix notation is represented by:
\begin{align}
\ket{x^{(i)}} \bra{x^{(i)}} = \begin{bmatrix} x_1^{(i)} x_1^{(i)} & x_1^{(i)} x_2^{(i)} & \cdots & x_1^{(i)} x_N^{(i)} \\
x_2^{(i)} x_1^{(i)} & x_2^{(i)} x_2^{(i)} & \cdots & x_2^{(i)} x_N^{(i)} \\
\vdots & \vdots & &\vdots \\
x_N^{(i)} x_1^{(i)} & x_N^{(i)} x_2^{(i)} & \cdots & x_N^{(i)} x_N^{(i)} \\ \end{bmatrix} \,.
\end{align}
Thus, the sum over training examples produces the following matrix:
\begin{align}
\frac{1}{M} \sum_{i=1}^M \ket{x^{(i)}} \bra{x^{(i)}} =\frac{1}{M} \begin{bmatrix} \sum_i x_1^{(i)} x_1^{(i)} & \sum_i x_1^{(i)} x_2^{(i)} & \cdots & \sum_i x_1^{(i)} x_N^{(i)} \\
\sum_i x_2^{(i)} x_1^{(i)} & \sum_i x_2^{(i)} x_2^{(i)} & \cdots & \sum_i x_2^{(i)} x_N^{(i)} \\
\vdots & \vdots & &\vdots \\
\sum_i x_N^{(i)} x_1^{(i)} & \sum_i x_N^{(i)} x_2^{(i)} & \cdots & \sum_i x_N^{(i)} x_N^{(i)} \\ \end{bmatrix}
\end{align}
which for demeaned data is equivalent to covariance matrix. The density matrix corresponding to correlation matrix could be obtained if the classical data is also standardized in the first step of algorithm. \\\\
\textbf{Exponential of density matrix} \\
Quantum phase estimation subroutine \textit{PhaseEstim} can be used to obtain eigenvectors and eigenvalues of the corresponding density matrix $\rho$. However, the quantum subroutine \textit{PhaseEstim} requires unitary matrix $U$ as an input, whereas generally density matrix $\rho$ does not meet that condition. In order to solve that problem we use mathematical law stating that $U=e^{iH}$ is unitary for any Hermitian matrix $H$. Obviously, density matrix $\rho$ is Hermitian by definition, so the only thing that remains is to take that density matrix and build the unitary gate $U=e^{i\rho}$. The knowledge of how physically exponentiate density matrix is not required to understand the algorithm, however the bulk of the algorithm consists of the ability to efficiently generate the exponent of an arbitrary density matrix $\rho$. The method is presented in Ref.~\cite{LloydPCA} and for the purpose of this algorithm it should be mentioned that the actual exponent generated $e^{-i\rho t}$ is with additional factor $t$ and the time complexity of generating this matrix is $\mathcal{O}(\log{N})$. \\\\
\textbf{Eigendecomposition of density matrix} \\
Having generated unitary matrix $U=e^{-i\rho t}$ we can apply quantum phase estimation subroutine \textit{PhaseEstim}. The additional factor $t$ in the exponent of unitary matrix would be helpful as in quantum phase estimation algorithm we use powers of unitary gate $U$. The eigenvectors of $\rho$ are also eigenvectors of $e^{-i\rho t}$ and the eigenvalues $\lambda$ of $\rho$ are just exponentiated $e^{-i\lambda t}$. The standard quantum phase estimation takes unitary gate $U$ as well as one of its eigenvector $\ket{\phi}$ and does the transformation:
\begin{align}
\ket{0}^{\otimes n}\ket{\phi} \rightarrow \ket{2^n\theta}\ket{\phi}
\end{align}
with phase $\theta$ in a control qubit allowing to calculate eigenvalue $\lambda=e^{2\pi i \theta}$ and $n$ denoting number of control qubits storing estimate of phase $\theta$. The qPCA algorithm uses slightly different phase estimation algorithm by applying unitary matrix $U=e^{-i\rho t}$ to density matrix $\rho$ instead of eigenvector. To see how it works, we firstly present the result of unitary matrix application on one of the pure state $\ket{x^{(i)}}$ from mixed state described by density matrix $\rho$:
\begin{align}
e^{-i\rho t}\ket{x^{(i)}} &= \sum_{j=1}^M e^{-i\lambda^{(j)}t} \ket{\phi^{(j)}} \braket{\phi^{(j)} \vert x^{(i)}} \nonumber \\
&=\sum_{j=1}^M e^{-i\lambda^{(j)}t} \braket{\phi^{(j)} \vert x^{(i)}} \ket{\phi^{(j)}} \nonumber \\
&=\sum_{j=1}^M c^{(ij)} \ket{\phi^{(j)}} 
\end{align}
with $c^{(ij)}=e^{-i\lambda^{(j)}t} \braket{\phi^{(j)} \vert x^{(i)}}$ Then the output of quantum phase estimation algorithm is a superposition:
\begin{align}
\ket{0}^{\otimes n}\ket{x^{(i)}} \rightarrow \sum_{j=1}^M c^{(ij)} \ket{\tilde{\lambda^{(j)}}}\ket{\phi^{(j)}} \,.
\end{align}
Due to the fact that we use unitary matrix $e^{-i\rho t}$ with eigenvalues $e^{-i\lambda^{(j)} t}$, the control state after \textit{PhaseEstim} subroutine equals $\tilde{\lambda^{(i)}}=2^n\frac{\lambda^{(i)}t}{2\pi}$ from which we can directly calculate eigenvalue $\lambda^{(i)}$. The output state can be also written in a density matrix representation as:
\begin{align}
\eta^{(i)}=\sum_{j=1}^M |c^{(ij)}|^2 \ket{\tilde{\lambda^{(j)}}}\bra{\tilde{\lambda^{(j)}}}\otimes \ket{\phi^{(j)}}\bra{\phi^{(j)}} \,.
\end{align}
In qPCA algorithm we apply the same logic, but instead of the pure state $\ket{x^{(i)}}$ we use mixed state $\rho$ described by density matrix $\frac{1}{M}\sum_{i=1}^M\ket{x^{(i)}} \bra{x^{(i)}}$. Weighting possible outputs $\eta^{(i)}$ of phase estimation algorithm with mixed state probabilities $\frac{1}{M}$ yields the final density matrix:
\begin{align}
\eta&=\sum_{j=1}^M \sum_{i=1}^M \frac{1}{M}|c^{(ij)}|^2 \ket{\tilde{\lambda^{(j)}}}\bra{\tilde{\lambda^{(j)}}}\otimes \ket{\phi^{(j)}}\bra{\phi^{(j)}} \nonumber \\
&=\sum_{j=1}^M \lambda^{(j)} \ket{\tilde{\lambda^{(j)}}}\bra{\tilde{\lambda^{(j)}}}\otimes \ket{\phi^{(j)}}\bra{\phi^{(j)}} \,. \label{eq:finalPCA}
\end{align}
The $\lambda^{(j)}$ coefficient is derived as follows:
\begin{align}
\sum_{i=1}^M \frac{1}{M}|c^{(ij)}|^2 &= \sum_{i=1}^M \frac{1}{M} e^{-i\lambda^{(j)}t} \braket{\phi^{(j)} \vert x^{(i)}} \braket{x^{(i)} \vert \phi^{(j)}} e^{i\lambda^{(j)}t} \nonumber \\
&= \bra{\phi^{(j)}} \left( \sum_{i=1}^M \frac{1}{M} \ket{x^{(i)}} \bra{x^{(i)}} \right) \ket{\phi^{(j)}} \nonumber \\
&= \bra{\phi^{(j)}} \rho \ket{\phi^{(j)}} \nonumber \\
&= \lambda^{(j)} \,.
\end{align}
To summarize, the difference is that in standard phase estimation algorithm an input is known eigenvector $\ket{\phi}$, whereas in \textit{PhaseEstim} subroutine used by qPCA an input $\rho$ is a mixed state with $M$ unknown eigenvectors $\ket{\phi^{(i)}}$. The time complexity of quantum phase estimation algorithm is comparable to exponentiating of density matrix thus does not increase the time complexity of algorithm. A quantum speed-up comes from the representation of classical data as qubits. \\\\
\textbf{Sampling} \\
Sampling from the final state derived in Eq.~(\ref{eq:finalPCA}) allows to reveal features of eigenvectors. Assume, that $R$ principal components sufficiently describe the variance of the data or equivalently the sum of eigenvalues corresponding to these principal components is close to $100\%$. The sampling of final state yields eigenvector $\ket{\phi^{(j)}}$ and corresponding eigenvalue $\lambda^{(j)}$ with probability $\lambda^{(j)}$. In other words, preparing $m$ copies of final state we would anticipate that the eigenvector $\phi^{(j)}$ will be sampled on average $m\lambda^{(j)}$ times. In case $m << R$ and providing that building the exponent of $\rho$ has $\mathcal{O}(\log{N})$ complexity, the time complexity of sampling is $\mathcal{O}(R\log{N})$. The sampled principal components in quantum states are then used to calculate scores or to reveal other features. \\\\
\textbf{Principal Components and Scores} \\
The following method is proposed to reveal the scores. The scores could be extracted from sampled principal component $\ket{\phi^{(j)}}$ by calculating the projection of an eigenvector on a training vector:
\begin{align}
\braket{x^{(i)} \vert \phi^{(j)}} &= \sum_{k=1}^N \sum_{m=1}^N x_k^{(i)} \phi_m^{(j)}\braket{k \vert m} \nonumber \\
&= \sum_{k=1}^N x_k^{(i)} \phi_k^{(j)} \nonumber \\ &= s_i^{(j)}
\end{align}
for each $i=\{1,\dots,M\}$ training vector encoded in quantum state and thus recreating score for $j$-th principal component:
\begin{align}
S^{(j)}=\left[s_1^{(j)}, s_2^{(j)},\dots,s_M^{(j)}\right]^T
\end{align}
The overlap $\braket{x^{(i)} \vert \phi^{(j)}}$ could be calculated by the quantum subroutine \textit{SwapTest}. The $R$ first scores corresponding to the highest eigenvalues represent the compressed data and can be used in another machine learning algorithm. Additionally, we can reveal additional features of sampled eigenvector $\ket{\phi^{(j)}}$ by measuring the expectation value $\braket{\phi^{(j)} \vert M \vert \phi^{(j)}}$ for an arbitrary observable $M$. The other machine learning algorithms could be accelerated if they can use these features in their set-up. Note that the qPCA algorithm is efficient in case $R<<N$, meaning that a relatively small number of principal components are required to explain variance of data. In other cases the time complexity of algorithm is similar to the classical one. 

\subsubsection{Summary}
The qPCA algorithm is based on the eigendecomposition of a density matrix $\rho$ and it is executed by \textit{PhaseEstim} subroutine. The sampling of a final state provides us principal components in quantum states from which we can calculate scores or reveal other features of eigenvectors. The time complexity of the algorithm consists of exponentiating density matrix $\rho$ in $\mathcal{O}(\log{N})$ and sampling a final state which results in the overall time complexity $\mathcal{O}(R\log{N})$. This is an exponential speedup comparing to $\mathcal{O}(N)$ time complexity for classical PCA. The qPCA speed-up applies only if a data could be explained by a relatively small number of principal components. The quantum speed-up is caused mainly by the representation of classical data as qubits. 

\subsection{Quantum Neural Networks}

\subsubsection{Preliminaries}
The classical neural networks algorithm uses the layers of connected neurons and selects the weights in each layer so that the cost function is minimized. The neurons are represented by activation functions such as sigmoid or softmax functions, which introduce the nonlinearity to the algorithm. The neural networks algorithm is mostly used in pattern recognition, classification problems and financial time series prediction in which they try to learn the non-linear dependencies between an input and output data. Currently, there is no specific quantum version of the algorithm, much work is concentrated on Hopfield neural networks, which are more close to neuroscience than to machine learning. In this section, the quantum feedforward neural networks (qNN) is presented as it is easier in the explanation. Moreover, the classical feedforward neural networks are very popular in machine learning applications. The qNN algorithm is based on the work of Wan et al. \cite{Wan} and is a straightforward generalization of classical feedforward neural network. The logic behind the algorithm is to turn classical components of a neural network to quantum components in a step by step manner. 

\subsubsection{Algorithm}
\noindent \textbf{Classical feedforward neural network} \\
We start with classical feedforward neural network. For simplicity, we assume that the input vector is only two dimensional $N=2$, characterized by classification label $y$ and there is only one hidden layer with one neuron. This could be represented by the graph:
\begin{figure}[H]
\centering
\includegraphics[width=0.5\textwidth]{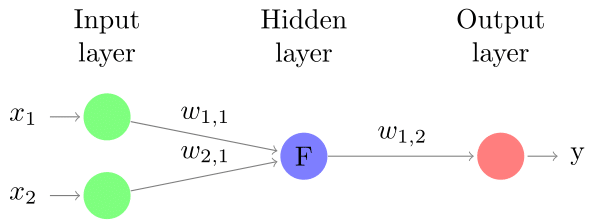}
\caption{Classical feedforward neural network}
\label{fig:NN-1}
\end{figure} \noindent
with $w_{i,l}$ indicating $i$\textit{-th input to layer} $l$ weight and the activation function denoted by $F$. Immediately, two problems arise with regard to quantum computations:
\begin{itemize}
\item The quantum operations need to be reversible and unitary. The classical activation function $F$ is not reversible as it squeezes the two dimensional vector to a single number $y$,
\item The quantum states multiplied by weights would not have any practical implications as states are always normalized. 
\end{itemize}
\textbf{Reversible feedforward neural network} \\
In order to fix irreversibility of classical neural networks, an extra variable denoted by $0$ is added to the input vector and the activation function $F$ is changed to output $x_1$, $x_2$ and $y$. To summarize, the function $F$ has three inputs and three outputs making the operation reversible. The graph presents the approach:
\begin{figure}[H]
\centering
\includegraphics[width=0.5\textwidth]{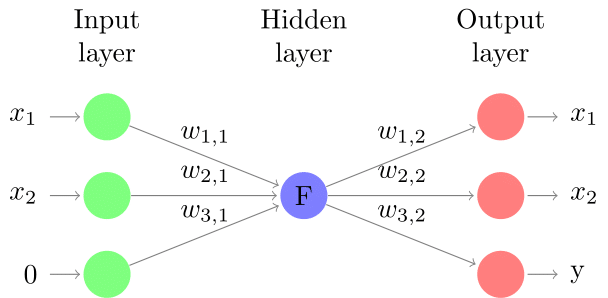}
\caption{Reversible classical feedforward neural network}
\label{fig:NN2-1}
\end{figure} \noindent
\textbf{Representation of classical data in quantum states} \\
We would like to represent classical $2$-dimensional feature vector $x=[x_1,x_2]^T$ together with label $y$ as quantum states. It can be achieved by representing classical data as binary strings:
\begin{align}
x_1 &\rightarrow \begin{bmatrix} a_{1,1} & a_{1,2} & \dots & a_{1,k} \end{bmatrix}^T \nonumber \\
x_2 	&\rightarrow \begin{bmatrix} a_{2,1} & a_{2,2} & \dots & a_{2,k} \end{bmatrix}^T \nonumber \\
y 	&\rightarrow \begin{bmatrix} b_1 & b_2 & \dots & b_m \end{bmatrix}^T
\end{align}
with $a_{i,j}=\{0,1\}$ for $i=1,2$ and $j=1\dots,k$ as well as $b_p=\{0,1\}$ for $p=1\dots,m$. We assumed that each entry of feature vector could be expressed in $k$ bits and label could be represented by $m$ bits. The binary strings can be then directly translated into $k$ qubit states for feature vector entries and $m$ qubit state for label:
\begin{align}
\ket{x_1}  	&= \ket{a_{1,1}} \otimes \ket{a_{1,2}} \otimes \dots \otimes \ket{a_{1,k}} \nonumber \\
\ket{x_2}  	&=\ket{a_{2,1}} \otimes \ket{a_{2,2}} \otimes \dots \otimes \ket{a_{2,k}} \nonumber \\
\ket{y}  		&=\ket{b_1} \otimes \ket{b_2} \otimes \dots \otimes \ket{b_m} 
\end{align}
Different ways could be chosen to represent a classical data in quantum states and the efficiency of the algorithm could be different for each method applied. \\\\
\textbf{Quantum feedforward neural network} \\
The last step is to change reversible function $F$ into an unitary quantum gate $U$, which inputs and outputs are quantum states. The graph below illustrates the final approach:
\begin{figure}[H]
\centering
\includegraphics[width=0.5\textwidth]{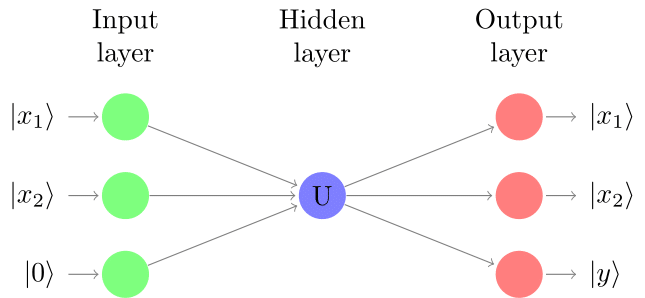}
\caption{Quantum feedforward neural network}
\label{fig:NN3-1}
\end{figure} \noindent
An arbitrary unitary matrix acting on $n$ qubits can be represented as:
\begin{align}
U=\exp{\left[i\left(\sum_{k_1,\dots,k_n=0,\dots,0}^{3,\dots,3}\alpha_{k_1,\dots,k_n}(\sigma_{k_1}\otimes\dots\otimes\sigma_{k_n})\right)\right]}
\end{align}
where $\sigma_k$ are Pauli matrices for $k=\{1,2,3\}$ and $\sigma_0$ is just 2x2 identity matrix. The parameters $\alpha_{k_1,\dots,k_n}$ are trainable parameters which can be obtained minimizing the cost function. This is different approach in comparison to classical  neural networks as the activation functions (instead of numeric weights) represented by unitary matrices  are trainable. The representation can be easily generalized to unitary matrix acting on any number of qubits. \\\\
\textbf{Minimization of cost function} \\
The cost function $C$ is proposed to be a sum of overlays over $M$ training examples (multiplied by $-1$ to reach solution by minimization). This is written as:
\begin{align}
C=-\sum_{j=1}^M\braket{y_{model}^{(j)} \vert y^{(j)}} \,.
\end{align}
The modeled $y_{model}^{(j)}$ is dependent on unitary matrix parameters $\alpha_{k_1,\dots,k_n}$, thus in each optimization iteration the problem reduces to selection of proper values $\alpha_{k_1,\dots,k_n}$, build a new quantum gate $U$ and calculate the value of cost function by retrieving the overlay $\braket{y_{model}^{(j)} \vert y^{(j)}}$ for each training example $j$. The gradient descent can be chosen as minimization procedure, updating the parameters $\alpha_{k_1,\dots,k_n}$ as follows:
\begin{align}
\delta \alpha_{k_1,\dots,k_n}= -\eta \frac{\partial C}{\partial \alpha_{k_1,\dots,k_n}}
\end{align}
with learning parameter $\eta$ controlling how much the coefficients are changing in each update. Obviously, the discretized version of gradient descent should be used. Alternatively, the paper \cite{Wan} defines the cost function as:
\begin{align}
C=\sum_{j=1}^M \sum_{i=1}^3 f_{i,j} (\langle \sigma_i \rangle_{model}-\langle \sigma_i \rangle)^2
\end{align}
where $f_{i,j}$ are non-negative real parameters and $\langle \sigma_i \rangle$ are the expectation values of Pauli matrices on individual outputs $y^{(j)}$. The Pauli matrices $\sigma_1$, $ \sigma_2$, $\sigma_3$ are observables of particle spin in $x$, $y$ and $z$ directions, respectively. Thus, the cost function expresses the difference between modeled labels and actual labels by measuring the observables on states $y_{model}^{(j)}$ and $y^{(j)}$ and comparing the average results. The more modeled labels and actual labels differs, the bigger will be the difference between expectation values, thus it is a valid definition of cost function. 

\subsubsection{Summary}
The benefits of quantum neural networks in comparison to the classical algorithm are not yet specified. A potential speed-up can come from a superposition, however the assessment of unitary matrix learning speed is difficult. The quantum algorithm presented is, in fact, classical in terms of learning parameters, so possible further improvement can be achieved by representing $\alpha_{k_1,\dots,k_n}$ parameters in a quantum state allowing for superposition. There are also concerns how the unitary matrix learning will be implemented experimentally. Nevertheless, the quantum neural networks algorithm is an interesting and promising field of quantum machine learning algorithms that is currently under development.

\section{Conclusions}
The work presented the key elements required to understand the quantum machine learning algorithms. The basic quantum theory, quantum computation and quantum algorithms formed a proper basis for the introduction of quantum machine learning algorithms. The author of this paper tries to simplify and efficiently pass the knowledge by grouping quantum algorithms into quantum subroutines and explaining them in a step-by-step manner with examples. The paper focuses on selected supervised and unsupervised machine learning algorithms and carefully describes their quantum versions. Nevertheless, the paper is by no means the complete review, further quantum algorithms such as quantum decision tree \cite{Lu}, quantum Bayesian methods \cite{gut,sasaki} and quantum associative memory \cite{trugen} exist in the literature. Moreover, the quantum annealing techniques implemented by adiabatic quantum computing are developed \cite{Lloyd,Wittek} and can be used as an alternative to quantum circuit approach. Most of the algorithms presented in the paper provided a significant speed-up in comparison to their classical counterparts, making it a promising field of quantum computation that can enhance the area of machine learning. On the other hand, one should remember that quantum circuit algorithms presented in the paper are sensitive to a decoherence, thus any noise present in a quantum system may destroy the benefits of quantum computing. Although the assessment of time complexities is conducted, the exact execution time is not certainly known until the algorithms would be run on a quantum computer. To sum up, even though there is much work required in theory and implementation, the quantum machine learning algorithms may solve problems that today's data scientists and company's management are facing in terms of time and calculation resources. Hopefully, the paper is just another step to accomplish this goal by making the knowledge about quantum algorithms more accessible. 

\bibliographystyle{plain}

\begin{thebibliography}{99}

\bibitem{MH}
M. Hilbert and P. L{\'o}pez,
\newblock The world's technological capacity to store, communicate, and compute information,
\newblock {\em Science}, 332(6025) (2011) pp.~60--65.

\bibitem{IBM}
\newblock IBM Builds Its Most Powerful Universal Quantum Computing Processors, \\
\newblock {\em https://www-03.ibm.com/press/us/en/pressrelease/52403.wss} (2017).

\bibitem{Broglie}
L. de Broglie,
\newblock XXXV. A tentative theory of light quanta,
\newblock {\em Phil. Mag} 47(278) (1924) pp.~446--458.

\bibitem{Davisson}
C. J. Davisson, L. H. Germer, 
\newblock Reflection of Electrons by a Crystal of Nickel,
\newblock {\em Proceedings of the National Academy of Sciences of the United States of America} 14(4) (1928) pp.~317--322.

\bibitem{EinsteinPhoto}
A. Einstein, 
\newblock Concerning an Heuristic Point of View Toward the Emission and Transformation of Light,
\newblock {\em Ann. Phys.} 17 (1905) 132.

\bibitem{Bohr}
M. Born, 
\newblock {\em The Born Einstein Letters}
\newblock  (Walker and Company, New York, 1971).

\bibitem{Griffiths}
D. J. Griffiths,
\newblock {\em Introduction to Quantum Mechanics}
\newblock (Cambridge University Press, 2016).

\bibitem{Quirk}
\newblock Quirk,
\newblock {\em http://algassert.com/quirk}.

\bibitem{QuTiP}
J. R. Johansson, P. D. Nation, and F. Nori,
\newblock QuTiP 2: A Python framework for the dynamics of open quantum systems,
\newblock {\em Comp. Phys. Comm.} 184 (2013) 1234.

\bibitem{Grover}
L. K. Grover,
\newblock A fast quantum mechanical algorithm for database search,
\newblock {\em arXiv preprint quant-ph/9605043} (1996).

\bibitem{Bennett}
C. H. Bennett, E. Bernstei, G. Brassard, U. Vazirani,
\newblock The strengths and weaknesses of quantum computation,
\newblock {\em SIAM Journal on Computing}, 26(5) (1997) pp.~1510--1523.

\bibitem{Coles}
P. J. Coles et al. 
\newblock Quantum Algorithm Implementations for Beginners,
\newblock {\em arXiv:1804.03719} (2018).

\bibitem{Durr}
C. D\"urr and P. H{\o}yer,
\newblock A quantum algorithm for finding the minimum,
\newblock {\em arXiv preprint quant-ph/9607014} (1996).

\bibitem{Ahuja}
A. Ahuja and S. Kapoor,
\newblock A Quantum Algorithm for finding the Maximum,
\newblock {\em arXiv preprint quant-ph/9911082} (1999).

\bibitem{Baritompa}
W. P. Baritompa, D. W. Bulger and G. R. Wood,
\newblock Grover's Quantum Algorithm Applied to Global Optimization,
\newblock {\em SIAM Journal on Optimization}, 15(4) (2005) pp.~1170--1184

\bibitem{Farhi}
E. Farhi, J. Goldstone and S. Gutmann,
\newblock A Quantum Approximate Optimization Algorithm,
\newblock {\em arXiv preprint quant-ph/1411.4028} (2014).

\bibitem{Verdon}
G. Verdon, M. Broughton and J. Biamonte,
\newblock A quantum algorithm to train neural networks using low-depth circuits,
\newblock {\em arXiv preprint quant-ph/1712.05304} (2017).

\bibitem{Hales}
L. Hales and S. Hallgren,
\newblock An improved quantum Fourier transform algorithm and applications,
\newblock {\em Proceedings of the 41st Annual Symposium on Foundations of Computer Science}, (2000) pp.~515.

\bibitem{Clave}
R. Cleve, A. Ekert, C. Macchiavello and M. Mosca,
\newblock Quantum algorithms revisited,
\newblock {\em Proceedings of the Royal Society A: Mathematical, Physical and Engineering Sciences}, (1998) pp.~454.

\bibitem{Manning}
C. D. Manning, P. Raghavan and H. Schütze,
\newblock {\em Introduction to Information Retrieval}
\newblock (Cambridge University Press, New York, 2008).

\bibitem{aimur}
E. A{\"\i}meur, G. Brassard, and S. Gambs,
\newblock Machine learning in a quantum world,
\newblock {\em Advances in Artificial Intelligence}, Springer (2006) pp.~431-442.

\bibitem{Lloyd}
S. Lloyd, M. Mohseni, and P. Rebentrost,
\newblock Quantum algorithms for supervised and unsupervised machine learning,
\newblock {\em arXiv preprint 1307.0411} (2013).

\bibitem{Niemann}
P. Niemann, R. Datta, and R. Wille,
\newblock Logic Synthesis for Quantum State Generation,
\newblock {\em IEEE 46th International Symposium on Multiple-Valued Logic}, Springer (2016) pp.~247-252.

\bibitem{Vartiainen}
M. Möttönen, J. Vartiainen, V. Bergholm, and M. M. Salomaa,
\newblock Transformation of quantum states using uniformly controlled rotations,
\newblock {\em Quantum Information and Computation}, 5 (2005) pp.~467-473.

\bibitem{aimur2}
E. A{\"\i}meur, G. Brassard, and S. Gambs,
\newblock Quantum clustering algorithms,
\newblock {\em Proc. 24th international conference on machine learning}, (2007) pp.~1-8.

\bibitem{anguita}
D. Anguita, S. Ridella, F. Rivieccio and R. Zunino,
\newblock Quantum optimization for training support vector machines,
\newblock {\em Journal Neural Networks - 2003 Special issue: Advances in neural networks research}, 16(5-6) (2003) pp.~763--770.

\bibitem{rebentrost}
P. Rebentrost, M. Mohseni, and S. Lloyd,
\newblock Quantum support vector machine for big data classiﬁcation,
\newblock {\em arXiv preprint 1307.0471} (2013).

\bibitem{LloydPCA}
S. Lloyd, M. Mohseni, and P. Rebentrost,
\newblock Quantum principal component analysis,
\newblock {\em arXiv preprint 1307.0401} (2013).

\bibitem{Wan}
K. H. Wan, O. Dahlsten, H. Kristjánsson, R. Gardner and M.S. Kim,
\newblock Quantum generalisation of feedforward neural networks,
\newblock {\em arXiv preprint 1612.01045} (2016).

\bibitem{Lu}
S.Lu and S. Braunstein,
\newblock Quantum decision tree classifier,
\newblock {\em Quantum Information Processing}, 13(3) (2014) pp.~757--770.

\bibitem{gut}
M. Gu{\c{t}}{\u{a}} and W. Kot{\l}owski,
\newblock Quantum learning: asymptotically optimal classification of qubit states,
\newblock {\em New Journal of Physics}, 12(12) (2010) 123032.

\bibitem{sasaki}
M. Sasaki, A. Carlini, and R. Jozsa,
\newblock Quantum template matching,
\newblock {\em Physical Review A}, 64(2) (2001) 022317.

\bibitem{trugen}
C. Trugenberger,
\newblock Quantum pattern recognition,
\newblock {\em Quantum Information Processing}, 1(6) (2002) pp.~471--493.

\bibitem{Wittek}
P. Wittek,
\newblock {\em Quantum Machine Learning: What Quantum Computing Means to Data Mining}
\newblock  (Academic Press, 2014).

\end{thebibliography}

\end{document}